\documentclass[aps,prf,preprint,groupedaddress,amssymb,onecolumn]{revtex4-1}
\usepackage[utf8]{inputenc}

\usepackage{natbib}
\usepackage{outlines}
\usepackage{indentfirst}
\usepackage{amsmath}
\usepackage{xfrac}
\usepackage{subfigure}
\newcommand{\degree}{^{\circ}}

\usepackage[usenames,dvipsnames]{pstricks}
\usepackage{epstopdf}
\usepackage{pst-grad} 
\usepackage{pst-plot} 

\newsavebox{\astrutbox}
\sbox{\astrutbox}{\rule[-5pt]{0pt}{20pt}}

\usepackage{ulem}

\usepackage{color}

\newcommand\modif[1]{\textcolor{black}{{#1}}} 
\newcommand\thie[1]{\textcolor{black}{{#1}}} 

\begin{document}

\title{Confronting Grand Challenges in Environmental Fluid Mechanics}

\author{T. Dauxois$^1$, T. Peacock$^2$, P. Bauer$^3$, C.P. Caulfield$^{4,5}$, C. Cenedese$^6$, C. Gorl\'e$^7$, G. Haller$^8$, G.N. Ivey$^9$, 
 P.F. Linden$^5$, E.~Meiburg$^{10}$, N. Pinardi$^{11}$, 
  N.M. Vriend$^{4,5}$, A.W. Woods$^4$}

\affiliation{1. Univ de Lyon, ENS de Lyon, Univ Claude Bernard, CNRS, Laboratoire de Physique, F-69342 Lyon, France\\
2. Department of Mechanical Engineering, Massachusetts Institute of Technology, Cambridge, MA 02139, USA\\
3. European Centre for Medium-Range Weather Forecasts, Reading RG2 9AX, UK\\
4. BP Institute, University of Cambridge, Madingley Rise, Madingley Road, Cambridge CB3 0EZ, UK\\
5. Department of Applied Mathematics and Theoretical Physics, University of Cambridge, Cambridge, CB3 0WA, UK\\
6. Woods Hole Oceanographic Institution, Woods Hole, MA 02543, USA\\
7. Department of Civil and Environmental Engineering, Stanford University, Stanford, CA 94305, USA\\
8. Institute for Mechanical Systems, ETH Z\"urich, Leonhardstrasse 21, 8092 Zurich, Switzerland\\
9. Oceans Graduate School, University of Western Australia, Crawley, Western Australia 6009, Australia\\
10. Department of Mechanical Engineering, University of California at Santa Barbara, Santa Barbara, CA 93106, USA\\
11. Department of Physics and Astronomy, University of Bologna, Bologna, Italy}

\date{\today}

\begin{abstract}
Environmental fluid {mechanics} underlies a wealth of natural, industrial and, by extension, societal challenges. In the coming decades, as we strive towards a more sustainable planet, there are a wide range of  grand  challenge problems that need to be tackled, ranging from fundamental advances in understanding and modeling of stratified turbulence and consequent mixing, to applied studies of pollution transport in the ocean, atmosphere and urban environments. 
A workshop was organized in the Les Houches School of Physics in France in January 2019 with the objective of gathering leading figures in the field to produce a road map for the scientific community. Five subject areas were addressed: multiphase flow, stratified flow, ocean transport, atmospheric and urban transport, and weather and climate prediction. 
This article summarizes the discussions and  outcomes of the meeting, with the intent of providing a resource for the community going forward.  

\end{abstract}

\maketitle

\section{Introduction} 

\begin{figure}
  \centering \includegraphics[width=0.65\textwidth]{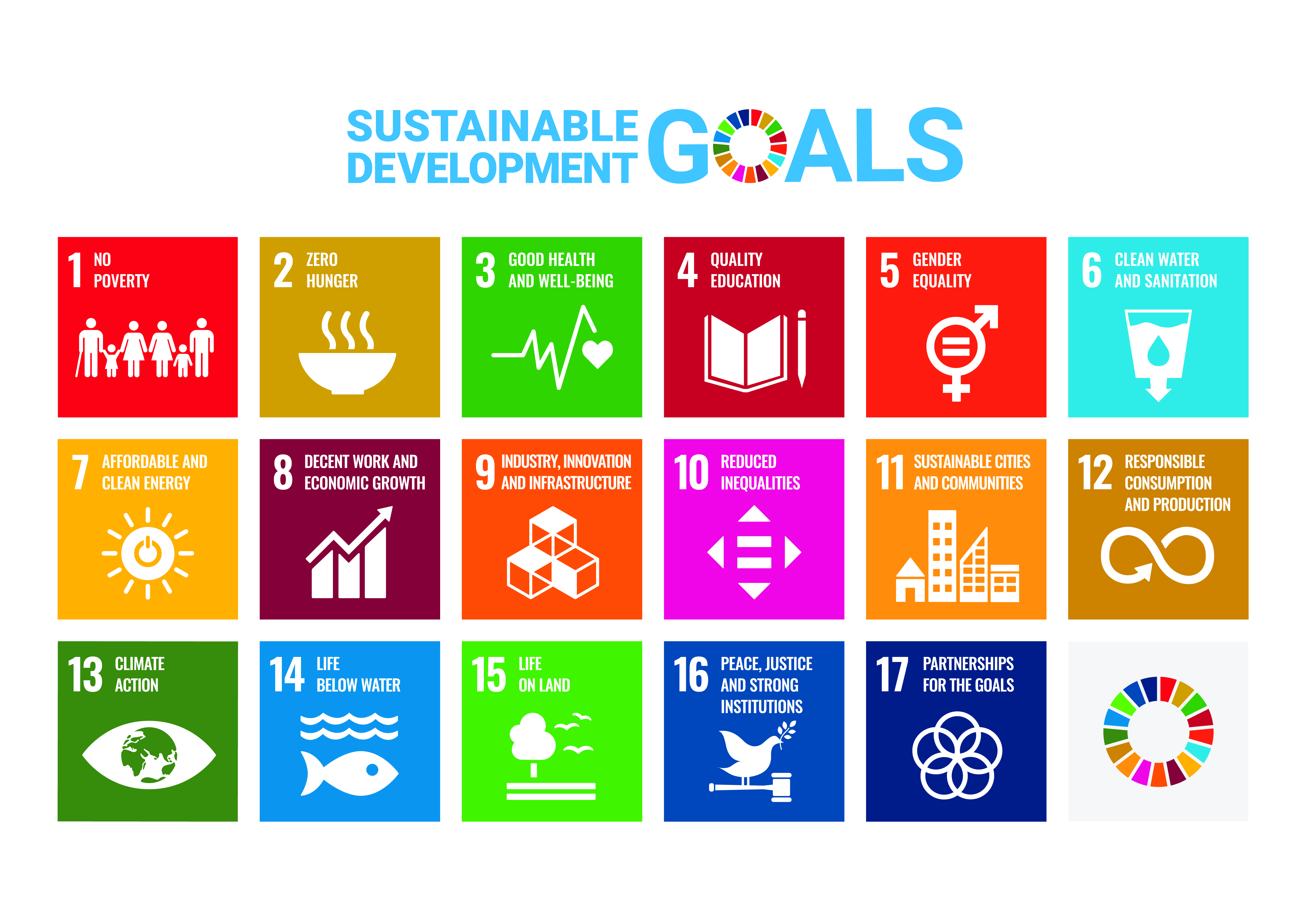}
\caption{
{The 2030 Agenda for Sustainable Development, adopted by all United Nations Member States in 2015, provides a shared blueprint for peace and prosperity for people and the planet, now and into the future. At its heart are the 17 Sustainable Development Goals (SDGs), which are an urgent call for action by all countries - developed and developing - in a global partnership. They recognize that ending poverty and other deprivations must go hand-in-hand with strategies that improve health and education, reduce inequality, and spur economic growth – all while tackling climate change and working to preserve our oceans and forests (https://www.un.org/sustainabledevelopment/). Of these, Environmental Fluid {Mechamics} plays a substantial role in achieving SDGs $\#$6, 7, 9, 11, 13 and 14.}}
  \label{fig:SDGs}
 \end{figure}

As the 21$^{st}$ century progresses, our planet faces numerous major environmental challenges, many of which are underpinned by environmental fluid {mechanics}. The modeling and monitoring of climate change and its consequences is perhaps the grandest of challenges, both to understand the system evolution and also to determine how some of the consequences may be mitigated {and adaptation plans devised. \modif{To help provide focus and guidance, researchers in environmental fluid mechanics can seek to support} many of the Sustainable Development Goals (SDGs) of the United Nations illustrated in Fig.~\ref{fig:SDGs}, in particular SDGs $\#$6, 7, 9, 11, 13 and 14, as outlined in this paper.}

The scientific community has a considerable capability to contribute to addressing environmental grand challenges {and achieving SDGs} by developing new understanding and innovating solutions. At a workshop~{\cite{TalksLesHouches}} at the Les Houches School of Physics in France in January 2019, therefore, a multifaceted group of seventy researchers convened to both identify and chart a way forward for grand challenges in environmental fluid {mechanics}. The outcomes of the resulting discussions are the focus of this article. Before delving into these grand challenges across a wide range of topics, however, it is initially worth reflecting on the scientific approaches available, and appreciating the broad spectrum of pressing scientific questions that lie within the realm of environmental fluid {mechanics}.

Field observations using innovative measurement systems gather valuable data on, and enable the description of, flow phenomena and processes. { Acquisition of high quality data, and interpretation of this data for developing and constraining models is at the heart of many of the grand challenges.}  A study on the mixing of North Atlantic Deep Water as it passes through the Tonga Trench in the deep Pacific Ocean provides new insight into the role of topography on abyssal mixing~\cite{REFClaudiaAndy}, a process that is key for quantification of the carbon and heat budget associated with the thermohaline circulation (see Sec.~\ref{Stratifiedflows}). The measurement of surface flows in the Gulf of Mexico using large arrays of low cost, degradable floats~\cite{tamay}, for example, identifies local points of convergence and highlights the importance of fronts in controlling surface transport, with clear relevance for the dispersal of oil spills (see Sec.~\ref{LagrangianCoherentStructure}).   Observations of the wind field and pollutant concentrations in buildings and urban areas have been shown~\cite{song_etal18,sousa2019} to be instrumental to the validation and improvement of computational models for these complex high Reynolds number flows.   {The recent global covid-19 pandemic has emphasized the importance of a Lagrangian understanding of air flows in sneezing and coughing and throughout buildings, in terms of the mixing pathways of airborne aerosols, bringing new challenges for the development of healthy and low energy building design~\cite{Bourouiba1,Bourouiba2,Mingotti,Andy1,Andy2}} (see Sec.~\ref{PredictingUrbanFlows}).

The development of analytical models complements field observations, with approaches ranging from dimensional analysis and the development of scaling laws to more complete theoretical models based on the appropriate fluid dynamical equations. Advances in theoretical modeling of environmental flows are very encouraging. Low order integral descriptions model the complex dynamics of mixing in turbulent jets and plumes, for example, and such models can be applied to natural ventilation flows through buildings~\cite{REFCarolinePaul}; often such flows are highly nonlinear and  exhibit multiple states, in a fashion analogous to the multiple states found in hydraulics~\cite{GladstoneWoods}, and the use of low order simplified models is ideal for identifying and interpreting such phenomena (see Sec.~\ref{PredictingUrbanFlows}). Research into salt fingering driven by double diffusive convection, which is key to understanding vertical mixing patterns in tropical oceans, has similarly been underpinned by fundamental understanding of scaling laws~\cite{Lohse}. Recently, classic models of sediment plumes have been used to underpin predictions of what might transpire from proposed deep-sea mining of minerals in the abyssal ocean~\cite{REFTom}.

Laboratory experimentation provides an invaluable means by which controlled, systematic and detailed studies can probe environmental flow phenomena and their evolution as the balance of forces changes. A key feature of laboratory experiments is their ability to access regimes that are challenging for analytical models, and to isolate and obtain detailed data on phenomena in a manner that is impractical for field studies. For example, laboratory experimentation is providing new insight into the important topic of microplastics transport in the ocean~\cite{REFDiBenidetto} and tsunami wave generation by a granular collapse~\cite{REFPhilippe}.

Numerical modeling comes to the fore for the study of geometrically, physically and dynamically complex scenarios,  producing extensive and detailed data sets that can be investigated via computer-based analysis methods. In regards to flow transport, for example, there have been significant advances using numerical methods to identify key Lagrangian coherent transport structures and track their evolution in time, with application to scenarios such as search-and-rescue operations at sea (see Sec.~\ref{LagrangianCoherentStructure}).
Advanced numerical techniques are now also available~\cite{Vowinckel2019} to simulate the evolution of suspension flows interacting with mobile sediment beds under increasingly realistic conditions (see Sec.~\ref{Multiphaseflow}).

A goal of large-scale computation is the accurate prediction of ocean and atmospheric weather patterns, and beyond that long-term climate patterns, for which the challenges are multifaceted~\cite{Dueben2018} (see Sec.~\ref{Prediction}). Approximations in the models include many sub-grid scale parameterizations of processes for which the physics is less well-understood, pertinent examples being stratified mixing (see Sec.~\ref{Stratifiedflows}) and convection dynamics (see Sec.~\ref{Prediction}). The approximations also include the incompleteness and error in observations used to condition the models, with the associated technical challenges of how best to assimilate data in such models; matters such as these necessitate an ensemble of model calculations to quantify uncertainty.  With increasing resolution of model systems (i.e. an increase of the number of grid cells), the science of data handling itself is becoming a limiting feature of large-scale computation.

{To recap, the main goal of this article is to highlight how environmental fluid {mechanics} can help answer critical questions as to the characterization of global climate change, develop solutions to mitigate this change following the SDGs directions and suggest adaptation strategies to climate change. Although not the central theme of this article, it is worth mentioning that environmental fluid {mechanics} will play an important role in the energy transition process (both supply and demand), which is likely to be one of the main endeavors of humankind in the present century and which is addressed in SDG $\#7$ (Access to affordable, reliable, sustainable and modern energy). The work on low energy buildings/natural ventilation (Sec.~\ref{PredictingUrbanFlows}) and the work on deep sea mining (Sec.~\ref{Multiphaseflow}) for cobalt (for batteries and hence electric vehicles) are good examples of the role of environmental fluid mechanics in the energy transition, which is a part of climate mitigation.}

The article is structured as follows. We begin in Sec.~\ref{Multiphaseflow} by describing the challenges related to multiphase flow, {covering topics such as water treatment (SDG $\#6$ Clean Water and Sanitation)} and the prediction of avalanches and volcanic eruptions {that pose a hazard to infrastructure and are hence relevant to SDG $\#9$ (Industry, Innovation and Infrastructure)}. We then move on to consider density stratified flows, which are relevant to scenarios such as vertical mixing in the deep ocean (see Sec.~\ref{Stratifiedflows}), {which is fundamental to a full understanding of the effects of climate change (SDG $\#13$ Climate Action)}. The transport of passive and active particles by environmental flows, the scenario relevant for pollutants transport in the ocean, is then the topic of  Sec.~\ref{LagrangianCoherentStructure} {and relevant for addressing SDG $\#13$ (Climate Action) and SDG $\#14$ (Life below water)}. This is followed by particular consideration of flows in urban environments, where the dispersal of pollutants and heat has a profound immediate impact on quality of life (see Sec.~\ref{PredictingUrbanFlows}) {and is the focus of SDG $\#11$ (Sustainable Cities and Communities)}. Then, weather and climate prediction{, relevant to addressing SDG $\#13$ (Climate Action),} are discussed in Sec.~\ref{Prediction}, with a viewpoint that the historic separation of these two fields is nearing an end because of the generic need for more realism in model physics. Finally, Sec.~\ref{Conclusion} concludes the article, outlining future directions for field experiments, theory, laboratory experimentation and numerical modeling in the field of environmental fluid {mechanics}. 

\section{Multiphase flow}
\label{Multiphaseflow}

\subsection{Introduction}

\begin{figure}[h]
\centering
\includegraphics[clip,width=0.8\textwidth]{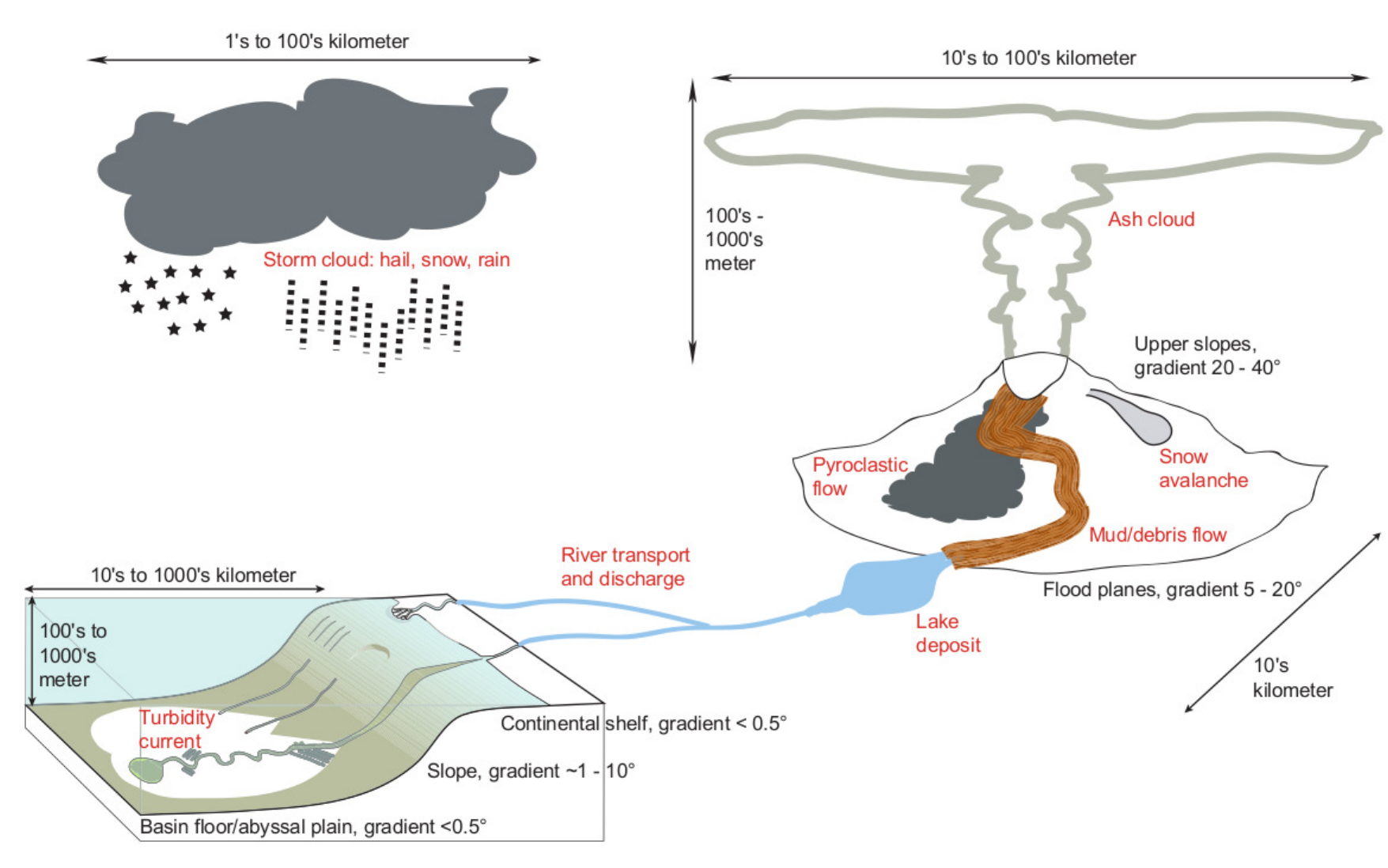}
\caption{{Multiphase flow processes in the environment at different scales: above us in the atmosphere, around us on land and below us in bodies of water.}}
\label{fig:ScalesMultiphase}
\end{figure}

Multiphase flow processes are ubiquitous in the environment, {as illustrated in Fig.~\ref{fig:ScalesMultiphase}}; above us, the dynamics of clouds are dominated by the interaction of air, water vapor, droplets and ice crystals, modulated by radiative heating and cooling. Around us, geophysical mass flows such as snow avalanches, mudslides, debris flows and volcanic eruptions present significant natural hazards. Below us, sediment transport processes in rivers, lakes and oceans affect the health of freshwater, estuarine and benthic ecosystems, as well as coastal and submarine engineering infrastructure. {Many environmental flow phenomena are man-made rather than natural} in origin, such as the transport of particular pollutants, the spreading of an oil spill {in the ocean}, or the generation of sediment-driven currents due to mining operations on the seafloor. The desire to better understand the drivers of climate change provides a major impetus for the rapidly growing research interest in environmental multiphase flows, as our limited understanding of such complex issues as the dynamics of clouds or the rate at which oceans absorb atmospheric {CO$_{2}$}, are among the largest uncertainties in existing climate models. {The feedback mechanisms between the changing climate and the evolution of glaciers and sea ice will greatly affect sea level rise and the security of freshwater supplies for a large fraction of the world's population. Similarly, the increasing intensity of wildfires, dust storms and dune migration due to climate effects poses a threat to people's livelihood in many dry regions of the world.}

A common feature shared by the above environmental multiphase flows is the enormous range of length scales to which they give rise, from droplets and clay particles of $O(10^{-6}$ m) to atmospheric weather systems and ocean currents of up to $O(10^6$ m). The resulting multiscale nature of the governing mechanisms renders the exploration of environmental multiphase flows by laboratory experiments, numerical simulations, field observations and remote sensing truly a 
{\it Grand Challenge}.

{Given the multitude of environmental multiphase flows, this section has to be selective by necessity, so that we will attempt to highlight only a few very active research areas of central importance in the context of the Sustainable Development Goals (SDGs) identified by the United Nations, especially with regard to climate change and its mitigation. The rapid progress in our understanding over the last couple of decades has been driven by improving diagnostic and modeling capabilities as a result of the availability of satellites, drones, and autonomous underwater vehicles, for example, as well as by more powerful computer hardware, computational algorithms, and other software tools.}

{In the following, we} distinguish between {\it dry} and {\it wet} environmental multiphase flows. In the former, interactions among particles dominate the overall dynamics while the interstitial fluid plays a relatively minor role. {This scenario applies, for example, to rock slides and certain types of snow avalanches that pose hazards to our infrastructure (SDG $\#9$ Industry, Innovation and Infrastructure). Similarly, issues of sand dune migration and desertification are of particular relevance in the context of promoting sustainable agriculture (SDG $\#2$ Zero Hunger), and sustainable use of terrestrial ecosystems (SDG $\#15$ Life on Land).}

{In wet multiphase flows,} on the other hand, viscous, pressure and buoyancy forces due to the presence of the fluid phase greatly influence the overall transport of mass, momentum and energy, so that they need to be properly accounted for when developing scaling laws and dynamical models. {Such conditions are encountered, for example, during the removal of particulate pollutants in water treatment plants (SDG $\#6$ Clean Water and Sanitation), or in the context of coastal erosion and the protection of infrastructure from the consequences of sea level rise. Climate modeling in particular (SDG $\#13$ Climate Action) gives rise to a host of interesting multiphase flow problems, for example associated with the dynamics of clouds, as will be discussed below. Further important examples of wet multiphase flows concern the transport of sediment, nutrients and microplastics in rivers and the coastal ocean (SDG $\#14$ Life below Water), or the dispersion of particulate pollutants in urban environments (SDG $\#11$ Sustainable Cities and Communities), a topic that is treated in more depth in {Sec.~V}.}

\subsection{Grand Challenges for dry flows}
{In such dry granular flows as rockslides or the migration of sand dunes,} the force distribution across the system is dominated by particle-particle interactions. In discrete particle models, the granular medium is characterized as a system of particles with trajectories determined by integrating Newton's equations of motion for each particle, resulting mathematically in a system of ordinary differential equations. The forces on an individual particle consist of an external gravitational force and contact forces resulting from particle-to-particle interactions depending on the selected contact model~\cite{Johnson1985}. Normal and tangential forces, including sliding and rolling resistance, are directly implemented as contact forces within the model. Discrete particle models retain the discrete nature of granular media, thus mimicking actual particle interactions closely, but are also limited by just generating point-data after every time-step, leading to computationally expensive simulations. Coarse-graining the output data is a necessary step to interpret the model results and to generate continuum fields.

In contrast, in continuum models the system loses direct access to particle-based properties as these are represented as local averages of position, velocity and stress fields. The fields are governed, and updated, through a system of partial differential equations prescribing the mass continuity and momentum balance of the system. The critical assumption here is to model the constitutive relation between kinematic (velocity) and dynamic (stress) fields accurately. Typical models for granular materials include the {$\mu(I)$-rheology~\cite{Jop2006}}, or the non-local cooperative~\cite{Kamrin2012} and gradient~\cite{Bouzid2013} models. In non-local models, it is assumed that the stress is not only a function of strain rate, but also depends on higher gradients of the velocity field. The particle size may be represented in constitutive models within the rheological description, but the exact scaling arguments are still an active topic of discussion.

An alternative to a full three-dimensional rheological model for granular materials is the depth-averaged model~\cite{Gray2014}. Here, the Saint-Venant shallow water equations are generalized, with one spatial dimension remaining in the governing equations. Although they are significantly easier and faster to implement numerically, one loses all information on the interior of the flow.

A different approach for studying dry granular flows is generating and using experimental data. A real-life experiment can show some truly unexpected behavior of particle dynamics; great examples of this are granular fingering~\cite{Pouliquen1997}, booming sand dunes~\cite{Hunt2010} and Faraday heaping~\cite{VanGerner2007}. The key to success is to represent all relevant physical processes and length-scales accurately in a scaled-down laboratory version of a full-scale environmental or industrial flow. Here, the use of effective non-dimensionalization is critical in order to identify dominant physical processes.

\begin{figure}[h]
\centering
\includegraphics[clip,width=0.8\textwidth]{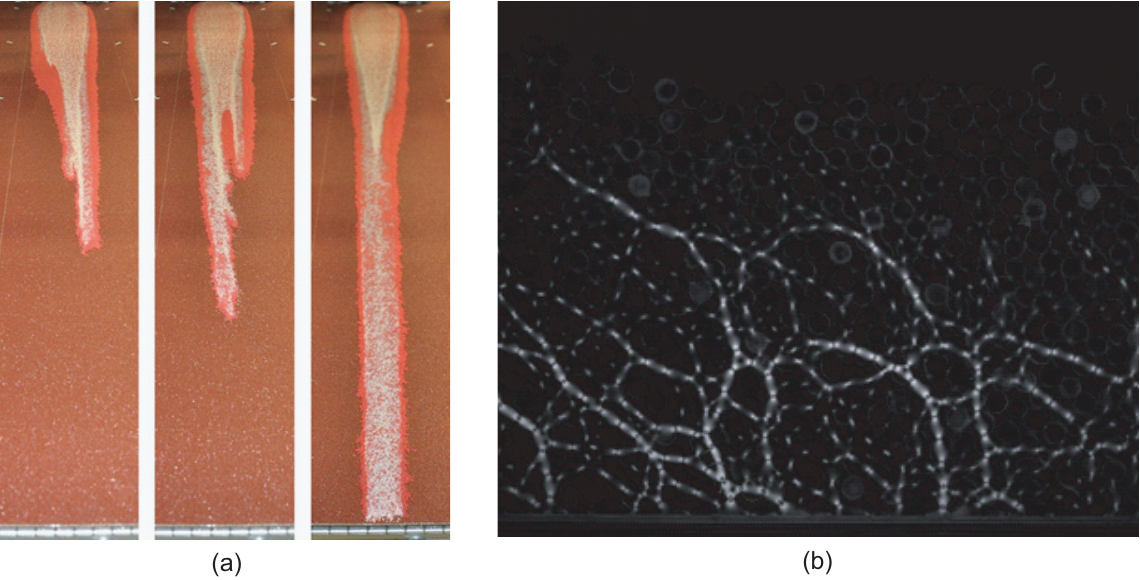}
\caption{{(a) Granular segregation leads to non-repeatable laboratory experiments. 
Release of 1~kg of a $50-50\%$ granular mixture of white (0.425 - 0.600mm) and red (1.0 - 1.3mm) glass ballotini from a 
2-meter long chute inclined at $25\degree$, 
showing different run-out lengths and levee structures. 
Experiments by Ms. Elze Porte. (b) Granular forces are transmitted in a non-homogeneous force-chain structure. Snapshot of a photoelastic 20-particle deep avalanche experiment down a two-dimensional channel with rough base. Experiments by Dr. Amalia Thomas.
}}
\label{fig:DryMultiphase}
\end{figure}

A wealth of experimental data on dry granular flows can be used to validate numerical simulations or test theoretical models. However, in order to do so effectively, data reduction needs to occur efficiently to deduce key properties and not get lost in big data sets. Experimental data may be limited in accuracy due to potentially small signal-to-noise ratios, or could be acquired with a larger than ideal spatial or temporal resolution. However, with the accessibility and affordability of high-speed cameras and advanced acquisition tools, the quality of experimental data improves year after year. A limitation in collecting experimental data is its granular nature; a small change in position of one single grain in the initial condition may create a completely different outcome, {as illustrated in Fig.~\ref{fig:DryMultiphase}(a)}. As a result, despite carefully controlled laboratory conditions, repeatability may be a concern and extensive data sets and even statistical analysis may be necessary.

A significant complication in acquiring experimental data is related to the opaqueness of dry granular materials, and the inability to look inside a dynamic flow. There are well-tested methods to probe immersed particulate flows, for example using refractive index-matched methods where a laser sheet and an interstitial fluid can reveal the dynamical behavior.  Dry particulate materials can be probed with X-Ray tomography, but this technique works only for quasi-steady set-ups, as it takes a significant time to acquire data~\cite{Weis2017}. Dry particulate flows in motion can be probed with Positron Emission Particle Tracking techniques~\cite{Parker2017}, but statistically significant data is difficult to obtain as there is only one tracer. The complication is that with all these techniques we only collect kinetic data on the velocity and position of individual particles, while we are not able to measure dynamic data revealing internal stresses and forces between particles.

Thomas and Vriend~\cite{Thomas2019} introduced the use of photoelastic analysis in gravity-driven intermediate flows to probe the rheology of {fast-moving} granular two-dimensional avalanches, {as illustrated in Fig.~\ref{fig:DryMultiphase}(b)}. Particle tracking and coarse-graining the point-data revealed both velocity and density profiles as a function of depth. Photoelastic analysis on the birefringent response, captured at sub-millisecond resolution, provides the full stress tensor with normal and shear stresses on each particle. Coarse-graining this data allows the calculation of the stress ratio and inertial number as a function of height, and tests the correlation between the shear rate and the force network fluctuations~\cite{Thomasetal2019}.

A fascinating example of dry particulate flows manifests itself ``out of our world'' in Martian dry gullies in the Avire Crater on Mars, where particulate material is present in an environment with no surface water, under low slopes~\cite{Dundas2017}. The high-resolution satellite images, which are collected at regular intervals in the High Resolution Imaging Science Experiment (HiRISE) by the Mars Orbiter Camera~\cite{McEwen2007}, provided  unprecedented images of erosion and transport of particulate material on low slopes in the Martian mid-latitudes. The creation and expansion of gullies coincides with seasonal {$CO_{2}$} frost, hence the physical process must be related to its presence.

\subsection{Grand Challenges for wet flows}
Turbidity currents {(underwater avalanches)} represent an excellent case study for reviewing some recent advances in our understanding of particle-laden flows, and for highlighting several open questions on which further progress is needed.

They represent the primary mechanism by which sediment is transported from shallow, coastal waters into the deep regions of the ocean~\cite{Meiburg2010}, and their size can be enormous. Often triggered by storms or earthquakes, a single large turbidity current can transport more than 100~km$^3$ of sediment, and it can travel over a distance exceeding 1,000~km, carving out deep channels on the seafloor. They are responsible for the loss of water storage capacity of reservoirs as a result of sedimentation, and they pose a threat to underwater engineering installations such as telecommunication cables and oil pipelines, {which renders them important in the context of the SDGs associated with safe drinking water supply {(SDG~$\#6$)} and sustainable infrastructure {(SDG $\#9$)}.} When triggered by submarine landslides near the coast, they can result in the formation of tsunamis. The sedimentary rock formed by turbidity current deposits represents a prime target for hydrocarbon exploration. {Turbidity currents are subject to the ocean transport mechanisms discussed in Sec.~IV, and they interact with the stratification of the ocean (cf. Sec.~III), which can give rise to such interesting phenomena as buoyancy reversal and lofting. At very large scales, their dynamics is furthermore affected by Earth's rotation.}

Far above the sediment bed, individual sediment grains are small, and their volume fraction is generally below $O$(1\%), so that particle/particle interactions are largely negligible. These dilute regions can be modeled by a continuum approach based on the Navier-Stokes Boussinesq equations, where the local density is a function of temperature, salinity and sediment concentration~\cite{Necker2002,Cantero2009}. The evolution of the sediment concentration field can be described by a convection-diffusion equation, where the sediment is assumed to move with the superposition of the fluid velocity and the Stokes settling velocity. Computational simulations based on this approach have provided substantial insight into the mixing and entrainment behavior of turbidity currents, along with their energetics. Investigations based on this dilute limit have furthermore shed light on the conditions under which particle-laden flows can give rise to double-diffusive instabilities. In particular, they have been able to clarify the competition between double-diffusive and Rayleigh-Taylor instabilities in the mixing region of buoyant river plumes and ambient salt water~\cite{Yu2014,Burns2015}. Very recently, linear stability analysis and nonlinear simulations based on the dilute approach have identified a novel, settling-driven instability mechanism in two-component flows, whose nonlinear growth can result in the formation of layers and staircases~\cite{Alsinan2017,Reali2017}.

Close to the sediment bed the dilute assumption no longer holds, as particle/particle interactions gain importance. These reduce the sedimentation rate of the particles through hindered settling. While some semi-empirical relationships for the effective settling rate in concentrated suspensions are available in the literature~\cite{Richardson1954,TeSlaa2015}, these were mostly obtained for conceptually simplified flow configurations, so that their reliability is questionable for sheared polydisperse mixtures of highly nonspherical particles consisting of heterogeneous materials. In addition, the particle/particle interactions render the fluid-particle mixture increasingly non-Newtonian, and there is considerable uncertainty with regard to its effective rheology. Recent years have seen progress through the development of the kinetic theory~\cite{Jenkins2002} and the $\mu(I)$-rheology~\cite{Boyer2011}, but their quantitative reliability has not yet been established for the complex conditions at the base of a large-scale turbidity current.

The situation is further complicated by deposition, erosion and resuspension. Early seminal work~\cite{Shields1936} quantified the threshold for erosion by considering the balance between gravitational and shear forces. Extensions of this work to date have been largely semi-empirical, and mostly consider idealized conditions, such as a dilute flow over a uniform sediment bed of monodisperse particles~\cite{Garcia1991}. Additional progress will have to be achieved in terms of quantifying erosion and deposition rates under complex flow conditions, before reliable predictions of field-scale turbidity currents become feasible. Advances in both computational and laboratory techniques offer promising opportunities in this regard, for example through further development of the 'smart sediment grains' technology~\cite{Frank2015}.

One important aspect that has received relatively little attention to date is the role of attractive interparticle forces, which can dominate for small sediment grains, such as mud, clay and silt. These cohesive effects prompt the primary grains to flocculate, and to form aggregates with larger settling velocities. Flocculation strongly affects such aspects as nutrient transport, and the rate at which organic matter is transported from the surface into the deeper layers of the ocean, with implications for modeling the global carbon cycle.

\begin{figure}[h]
\centering
\includegraphics[clip,width=0.8\textwidth]{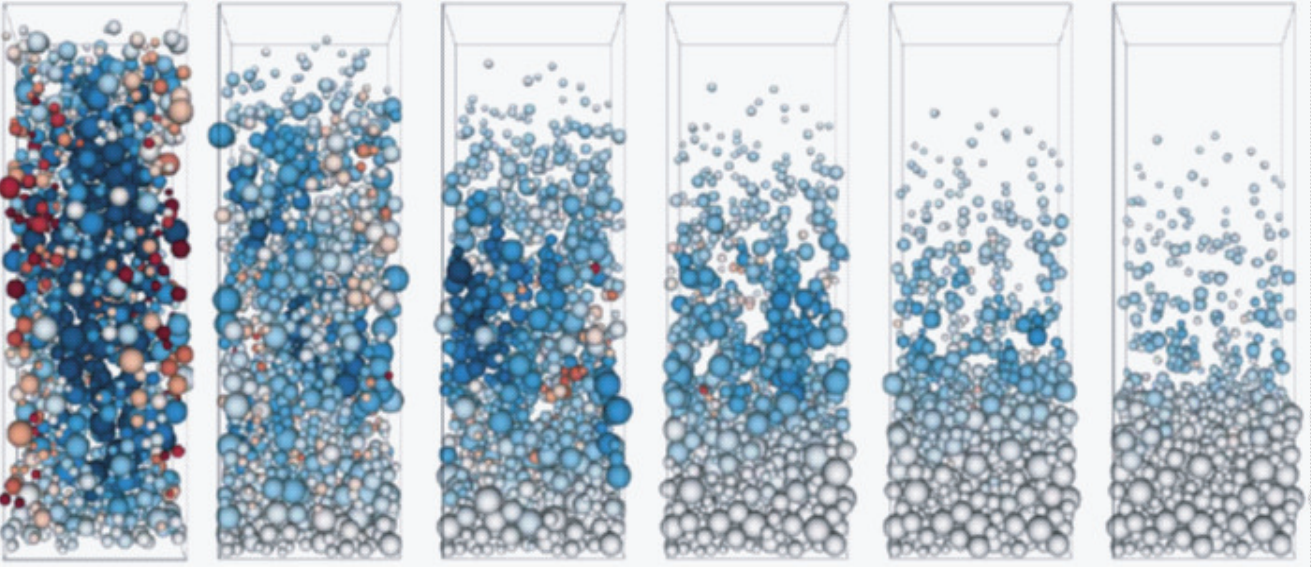}
\caption{{Particle configurations at different times during the settling process of cohesive sediment. The color reflects the vertical particle velocity. Adapted from \cite{Vowinckel2019}.}}
\label{fig:WetMultiphase}
\end{figure}

Recent years have seen significant advances through the advent of grain-resolving simulation approaches that allow for the tracking of thousands of interacting particles~\cite{Balachandar2010}. Frequently these numerical models are based on variations of the Immersed Boundary Method~\cite{Mittal2005}, which allows for the accurate and efficient tracking of moving interfaces within the framework of regular Cartesian grids. Similarly, more realistic collision models for particle-particle interactions~\cite{Biegert2017} have enhanced our ability to simulate the evolution of suspension flows interacting with mobile sediment beds under increasingly realistic conditions. Vowinckel {\it et al.} \cite{Vowinckel2019} have recently conducted the first grain-resolving simulations of cohesive sediment, in which they considered the sedimentation of 1,261 polydisperse particles, {as illustrated in Fig.~\ref{fig:WetMultiphase}}.

Multiphase environmental flows are often strongly affected by phase change. {An important case in point concerns the central importance of} condensation and evaporation for the evolution of atmospheric clouds~\cite{Shaw2003}. {By driving the global circulation and modulating the radiative and turbulent atmospheric transport of heat, mass and momentum in the presence of water phase changes, clouds represent a key element within the complex feedback loops that govern the dynamics of Earth's weather and climate (SDG $\#13$ Climate Action), cf. Sec.~VI. The dynamics of clouds, including their radiative properties and precipitation efficiency, are governed by a host of physical mechanisms that are active over a wide range of length scales, from the condensation, evaporation and collisional growth of individual droplets/ice particles at the $\mu$m-scale, via the formation of thermal and hydrodynamic instabilities at intermediate scales, to the turbulent transport of heat, mass and momentum at the km-scale and beyond, where interactions with larger-scale cloud systems and other phenomena occur. Our limited current understanding of cloud microphysics, and the associated lack of upscaling and parameterization tools for incorporating cloud dynamics into global climate models, represents a major source of uncertainty in the field of climate prediction. Nevertheless, recent advances in high-fidelity, large-scale computational simulation techniques, upscaling strategies, machine learning approaches, and experimental/observational capabilities provide opportunities for developing physics-based cloud models that can transform the field of climate prediction.}

A less well-known situation {of environmental multiphase flows with phase change} concerns the formation and precipitation of salt crystals in hypersaline lakes, such as the Dead Sea~\cite{Ouillon2019}. These processes are governed by the convective and diffusive transport of heat and salinity, as well as by the thermodynamic properties of brine near the saturation limit, and they can be strongly affected by gravity currents, double-diffusive instabilities and internal waves, among other features. The computational exploration of these phenomena is still in the very early stages.

While we can employ high-resolution computational approaches to investigate the microscale dynamics, the large range of scales requires suitable upscaling approaches to field scales. Developing such upscaling approaches to provide accurate predictions poses a significant challenge to the research community. Open source efforts such as the Community Surface Dynamics Modeling System (CSDMS, {\tt https://csdms.colorado.edu/wiki/Main$_{-}$Page}) can play an important role in this regard, as they try to couple models across different scales. There are numerous other interesting and highly relevant multiphase environmental flow processes that cannot be discussed within the limited space available here. Among the most fascinating problems are those involving ``active matter", such as the behavior of a swarm of insects~\cite{Kelley2013}, the contribution of plankton swarms to the mixing of the oceans~\cite{Houghton2018,Ouillon2019b}, or the flow of human crowds~\cite{Ouellette2019}. Yet another class of fascinating examples of multiphase flows in the environment involves capillary forces, such as in wet granular flows~\cite{Herminghaus2005}.

\subsection{Outlook}
{The study of environmental multiphase processes in the context of the Sustainable Development Goals is an exciting and vibrant field with new methods and techniques appearing in rapid progression. The available tools for fieldwork, laboratory experiments and numerical simulations are continuously improving in their capabilities. In fieldwork, drones and microsatellites are now deployed to obtain an unprecedented quality and quantity of field data, with details revealed which were previously unknown. In laboratory experiments, the spatial and temporal resolution which can be achieved in carefully-controlled conditions continues to improve with advancing technology. Innovation is a necessary tool to make steps forward to measure relevant physical properties and to allow the crossing of scales between real-life field observations and scaled-down laboratory analogues. The general strategy in numerical simulations is to explore the relevant physics at the microscale by creating more realistic computational models, and to combine those with upscaling tools to bridge the gap to larger scales. The large amount of experimental or numerical data that is generated in the study of particulate multiphase flow can now be post-processed by machine learning tools, to exploit our data progress and to enhance predictive capabilities.}

\section{Stratified flow}
\label{Stratifiedflows}

\subsection{Introduction}
Flows in the environment are typically characterised by
spatial and temporal variations in the fluid density, due for example
to variations in temperature  or composition, associated with
salinity, particle concentration, or other stratifying agent. 
Under appropriate  averaging (denoted
by an overline), \modif{much of} the atmosphere, the world's oceans and lakes are
statically stably stratified, with the  ``background'' or mean density
$\overline{\rho}$ decreasing upwards, \modif{although there are  also situations where this stable stratification is eroded (e.g. in the upper ``mixed'' layer of the ocean) or
even inverted to become statically unstable, such as in a ``convective'' atmospheric boundary layer.}
Such \modif{typical statically stable}
background density variations lead naturally to a  definition of the
``buoyancy frequency''~$N$, where
$ N^2 {\equiv} -({g}/{\overline{\rho}})({\partial
   \overline{\rho}}/{\partial z})$,
and $g$ is the acceleration due to gravity.  This buoyancy frequency
is the frequency of oscillation for a fluid parcel displaced
vertically within the background density profile, and bounds above the
possible frequencies of ``internal gravity
waves'' which are ubiquitous in the environment. Developing an understanding of
the mechanisms by which such waves are generated, propagate,  and ``break''
(thus nonlocally transferring momentum and energy \modif{and creating turbulence})  is an active area
of research~\cite{sarkar:2017}.

\modif{Of course, the effects of rotation are central to understanding the dynamics of the (generically) stratified fluid flows on earth. It is still very important to understand the behaviour of ``environmental'' flows, where the effects of rotation are assumed to be (largely) insignificant, not least because of the complex ways on which
  such relatively small-scale and fast flows can feed back on and nonlinearly affect larger scale
flows for which rotational effects may  not plausibly be ignored.} 

Even when the effect of rotation can be
discounted, \modif{the inevitably more ``modest'' research goal of} {\it in situ} observation and idealized modelling of such stratified flows is extremely
challenging, not only because of the vast range of scales that are observed but also due to
the generic appearance of spatio-temporally
intermittent turbulence.
The Grand Challenge
to the research community is thus to improve parameterization in
larger scale models 
of stratified  \modif{turbulent flows}, particularly the associated mixing
and transport effects, which
are fundamental to a full understanding of the effects of climate change (SDG $\#$13
Climate Action).
\modif{This parameterization is a key component in ocean circulation models used, for example, for environmental management and assessing the effects of climate change on ocean dynamics. It is widely acknowledged that this key ``building block'' remains an outstanding area
  of both controversy and uncertainty (see for example \cite{gregg:2018} for a  more detailed discussion of some of the central challenges).}
\modif{Mixing is important not only in large scale systems such as the oceans and the atmosphere. Smaller scale systems, such as catchments, lakes, water supply reservoirs and estuaries, are all closely connected to regions of human habitation. Quantifying  mixing in these water bodies  is key to the  predictive ability of aquatic ecosystem models (see for example \cite{hipsey:2020}, with direct application to  ensuring clean water and sanititation (SDG $\#$6), sustainable cities and communities (SDG $\#$11), and life below water (SGD $\#$14) - SDGs common with a number of other sections in this review.}

A key objective \modif{in all these applications}
is to parameterize how turbulent motions in a 
stratified fluid irreversibly mix the fluid, and thus transport heat
and other scalars vertically, or more precisely across density surfaces (and hence 
``diapycnally''). Attempts to parameterize such turbulent diapycnal 
transport is 
a very active area of research, using
both idealized ``academic'' studies of fundamental
fluid processes using laboratory experiments and (increasingly) high resolution
numerical simulations, and also {\it in situ} observation and measurement of
processes in  full-scale environmental flows.
It is very important to appreciate that
there are inevitable and substantial differences in the spatio-temporal resolution
and the quantity of data \modif{associated with a specific mixing event} obtainable from observation as compared to data from
simulation and laboratory experimentation.

A fundamental issue is then to ensure  synergistic communication
between these three classes ({\it i.e.} simulation/experimentation, observation and
parameterization) of research activity. This is proving, to put it mildly,
difficult. Perhaps the most straightforward way to understand this
difficulty is to appreciate that the progression from simulation
through observation to parameterization involves an inevitable
increase in complexity of the flow (in geometry, boundary conditions
and mean flow, for example) with a concomitant decrease
in the quantity and quality of available data. In particular,  there is
an
unnerving gap between the detailed descriptions available
from  simulations and laboratory experiments of idealized flows, and both
the available observations and parameterizations of the  systems of
interest. 
Nevertheless, recent developments in both modelling and observation
are starting to bridge these gaps suggesting that the research community
is
on the cusp of making major advances in constructing new and useful parameterizations
of turbulent mixing in stratified flows, an undoubted Grand
Challenge in environmental fluid dynamics.

\subsection{Grand Challenges for modeling}

The most basic  parameterization of mixing in stratified flows is the construction of a model for the 
(vertical) eddy diffusivity of density $K_\rho\equiv {\mathcal B}/N^2$,
a closure relating an appropriately defined vertical
buoyancy flux ${\mathcal B}$ to $N$. There are  two classic approaches  to the parameterization of
$K_\rho$, arising either from the equation for turbulent
kinetic energy or from the equation for density variance.
In an
exceptionally important and influential paper \cite{osborn:1980}, Osborn
 postulated in a statistically steady
state
that  ${\cal B}=\Gamma \epsilon$,
where $\epsilon$ is the  dissipation rate of turbulent kinetic energy,
such that the turbulent flux coefficient (sometimes called the
``mixing efficiency'') $\Gamma \leq 0.2$ (the inequality is very commonly 
ignored and instead replaced by an equality, see e.g.  \cite{waterhouse:2014}).
This appealing assumption greatly simplifies the problem, but assumes there is always
a fixed partitioning of turbulent kinetic energy between the  two ``sinks''
\modif{associated with irreversibly increases in the potential energy and viscous dissipation.}
Alternatively, Osborn and Cox~\cite{osborn:1972} postulated that ${\mathcal B}$ should be in balance
with the rate of destruction of the buoyancy variance $\chi$ which, distinctly different from the Osborn model,  requires no assumption about the kinetic energy
balance within the flow.

There are  a wide range of as yet un-resolved issues with
these two parameterizations that lie at the heart of much
of the analysis of observations~\cite{waterhouse:2014}, proposed improved parameterizations~\cite{salehipour:2016,mashayek:2017} and indeed
larger-scale models. We highlight a (small) subset of these questions below, which
were discussed during the workshop (further discussion of
the fundamental issues facing mixing parameterization
can be found in the reviews~\cite{ivey:2008,gregg:2018}). \modif{We then finish with a brief outlook on some of the future challenges around mixing in stratified flows.} 

\subsubsection{Time Dependence and Irreversibility}
\begin{figure}[t]
\includegraphics[width=\textwidth]{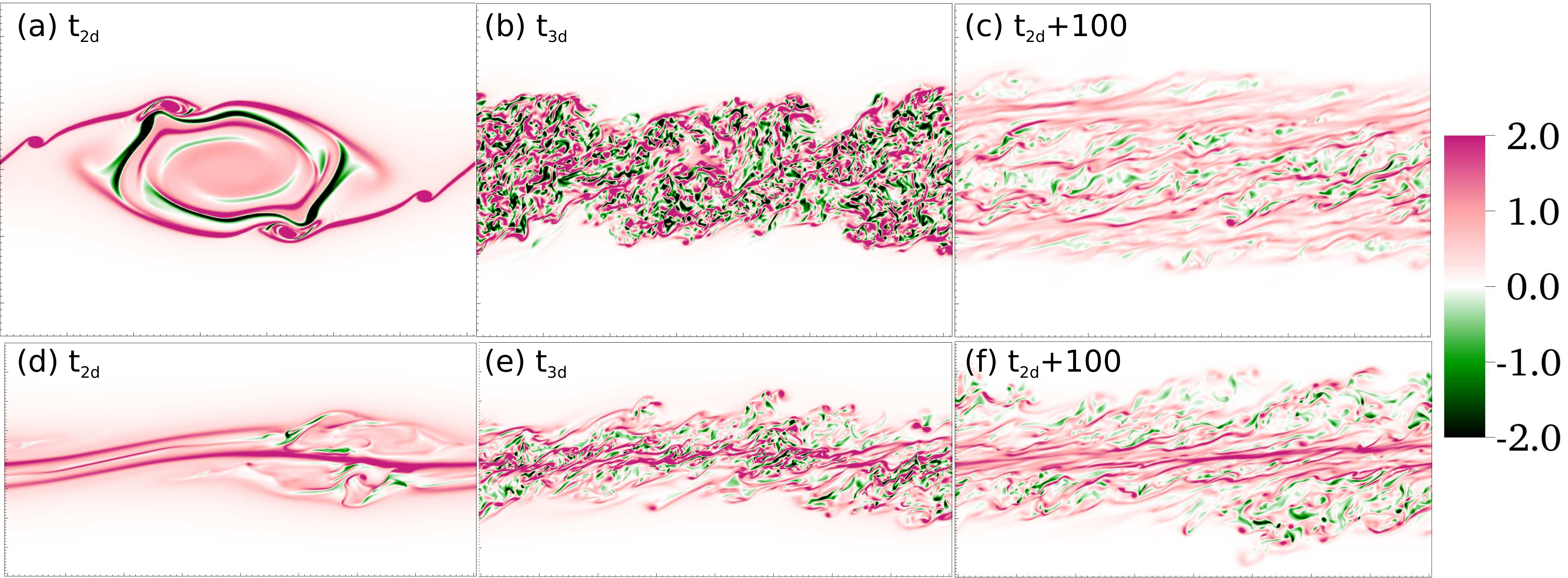}
\caption{Contours of the spanwise vorticity for simulations with the same
  initial Reynolds number and bulk Richardson number but
  (a-c) prone to primary KHI where the initial shear layer depth and density interface
  depth are equal; and
  (d-f) prone to primary HWI where the density interface is significantly ``sharper'' than the shear layer depth; at $t=t_{2d}$ (scaled with
  advective time units $d_0/U_0$),
  when spanwise-averaged TKE is maximum; $t=t_{3d}$ when
  three-dimensional perturbations are maximum; $t=t_{2d}+100$.  Note `overturning' by
  the primary KHI and `scouring' by the HWI.
Used
  with permission from \cite{salehipour:2016b}, copyright CUP,  all
  rights reserved.}
\label{fig:kh}
\end{figure}

Typical real-world
mixing events are inherently time-dependent and transient, and it is
not even clear what is the appropriate way to define the buoyancy frequency~\cite{arthur:2017} when there is vigorous
turbulence, associated with statically unstable overturning regions.
Indeed, in real flows it is not even required that the buoyancy
frequency is always positive and this can be seen, for example, in classical ``Kelvin-Helmholtz billow'' shear instabilities, denoted KHI~\cite{smyth:2001,mashayek:2013,salehipour:2015}.
\modif{The  time evolution of this instability is shown in the upper panels of Fig.~\ref{fig:kh}.}

Typically, the irreversible mixing rates constructed
using the ``background potential energy'' formalism
\cite{winters:1995}, has been used to construct
(irreversible) estimates for~$\Gamma$ within the Osborn model, although
such time-dependent mixing events clearly violate the underlying assumptions
of that model~\cite{mashayek:2013}. Indeed, 
through a careful comparison of different expressions,
\cite{salehipour:2015} demonstrated that an ``irreversible'' Osborn-Cox model
was more accurate than the Osborn model with fixed $\Gamma$ 
in capturing the actual mixing in a time-dependent Kelvin-Helmholtz
mixing event.
Interestingly, there is also recent observational evidence 
\cite{ivey:2018} that using  the Osborn-Cox model leads to better
estimates
of irreversible mixing, at least
in energetic flows where the turbulence
is strong relative to the stabilising effects of 
stratification.
It is plausible
that
the Osborn-Cox model, based as it is on properties
of the density field, is likely to be a better model for mixing than
the Osborn model, which inevitably has to ``pass through'' intermediate
modelling
assumptions
relating kinetic energy dissipation processes to mixing. This has
significant
implications both for future areas of focus in numerical simulation,
and also in terms of observational measurement where  the use of
recently-developed, robust methods~\cite{bluteau:2017} for   direct measurement of $\chi$ should be prioritised if at all possible. 

Furthermore, it is
certainly
not settled that  KHI-induced turbulent  mixing is a \modif{robust conceptual model} for
stratified
turbulent mixing in general, not least because
the relatively large-scale primary
overturning leaves an imprint throughout the entire subsequent
(relatively
short-lived) ``flaring''
life cycle, as discussed by \cite{mashayek:2017_kh}.
Even accepting that shear instability initial value problem
simulations lead to turbulence with the appropriate characteristics,
it is possible that instabilities which ``burn'' through longer mixing life cycles may be better \modif{conceptual models} for
environmentally-relevant stratified mixing events. 
\cite{salehipour:2016b, salehipour:2018}
has investigated the  turbulent mixing behaviour triggered by ``Holmboe wave
instabilities'' (HWI)
characterised by counter-propagating cusped waves, and associated with
relatively ``sharp'' density interfaces embedded within relatively
extended shear layers.
\modif{The time evolution of these instabilities is shown in the lower panels of Fig.~\ref{fig:kh}.}
These instabilities do not ``overturn'',  but
rather  ``scour''  the interface,  a mixing characterized by mixing
coefficients $\Gamma \simeq 0.2$, perhaps
fortuitously, similar to the canonical value of the Osborn model.

Even though such flows can exhibit vigorous turbulent motions above and below the density
interface, the notional spatio-temporally varying gradient
Richardson number $\overline{Ri}(z,t){\equiv}[(-g/{\bar{\rho}}) \partial
     \overline{\rho}/\partial z] /  [ \partial
       \overline{U}  /\partial z  ]^2$
has a probability density function (for varying $z$ and~$t$) strongly
peaked around $1/4$.
The specific value of $1/4$ has 
great significance in stratified shear flows, as 
\cite{miles:1961,howard:1961}
established that a necessary condition for linear normal-mode 
instability of an inviscid steady parallel stratified shear flow is 
that 
the Richardson number $\overline{Ri} < 1/4$ somewhere within the flow.
\cite{thorpe:2009} conjectures
that this specific value is still relevant to the dynamics
of turbulent flows where the ``background'' profiles
defining the Richardson number are notional
constructs from some averaging process of a time-dependent
flow, (which naturally does not satisfy the underlying assumptions of
the Miles-Howard theorem) with the intermittent
onset of instabilities maintaining the flow in a ``marginally stable''
state. The $\overline{Ri}$ data from these HWI simulations are suggestive that
there may indeed be a way in which turbulent flows
adjust towards such a marginally stable state, perhaps associated
with the concept of ``self-organised criticality''
\cite{smyth:2019}. Such works are suggestive of an as-yet unexplained
robustness in the relevance of linear stability analyses to turbulent
flows. 

\subsubsection{Forcing and Parameter Dependence}

Freely-evolving
shear-induced turbulence is by no means the only way in which
stratified mixing may be induced, and it is also an open question of
significant
interest whether explicitly forced,  \modif{unsheared or even convective} flows are
qualitatively different. 
Indeed, as discussed by~\cite{ivey:1991}, and more
recently by \cite{maffioli:2016} and \cite{garanaik:2019},
a perhaps more appropriate parameter to
describe the mixing properties of stratified turbulence is the
turbulent Froude number $Fr_T={\epsilon}/(N {\cal K})$, 
as it seems reasonable that the actual intensity ${\cal K}$ of the
turbulence
should be important, as well as its dissipation rate. As a parameter, $Fr_T$ also
has the attraction that it does not  rely on a  background shear.
This
point leads to perhaps {\it the} key open question: is it possible
(or useful) to attempt to identify generic properties of mixing
induced
by stratified turbulence, or is it always necessary to identify the
underlying forcing or driving mechanism ({\it e.g.} shear instabilities,
convective processes, topography etc) triggering the ensuing
irreversible mixing?  This is by no means settled
among the fluid dynamical community, and certainly deserves further
consideration. 

\subsubsection{Length scales}

\modif{Irrespective of the driving mechanism, the} various nondimensional parameters can also be interpreted as ratios of
key length scales.
For example, the buoyancy Reynolds number 
$Re_b \equiv \epsilon/(\nu N^2)= (L_O/L_K)^{4/3}$,
where $L_O\equiv (\epsilon/N^3)^{1/2}$ is the Ozmidov scale,
which 
may be interpreted as the largest vertical scale that is mainly unaffected 
by buoyancy effects, and 
$L_K$  is the Kolmogorov microscale.
Expressed in this way, it is thus apparent for
there to be any possibility of an inertial range of isotropic
turbulence, (characterised by scales $\ell_i$ both very much larger than
the dissipation scale $L_K$ and very much smaller than the energy
injection
scale), it is necessary that $Re_b \gg 1$.
Also, for the mixing ``grand challenge'', the parameter $Re_b$ is very
important,
not least because oceanographic flows are often characterised by very large
values of $Re_b$ \modif{\cite{gargett:1984}}. Furthermore, 
$K_\rho\equiv \nu \Gamma Re_b$, and there is ongoing controversy as to what (if any) \modif{is the dependence of $\Gamma$ on
 $\overline{Ri}$, $Re_b$ and $Fr_T$}~\cite{ivey:1991,shih:2005,maffioli:2016,salehipour:2016,mashayek:2017,gregg:2018,monismith:2018,garanaik:2019,portwood:2019}.

A further length scale which has attracted much interest is the
so-called ``Thorpe'' scale~$L_T$.
In particular, the ratio $R_{OT}= L_O/L_T$ has been
proposed both as a measure of the ``age''\modif{\cite{dillon:1982}} of a specific patch of
turbulence, and also as a way to infer $\epsilon$, and hence
mixing, using (for example) the Osborn model with fixed $\Gamma$. Unfortunately, it is 
clear that there are significant issues with this approach both from
observational data and numerical simulation (e.g.
\cite{mater:2015} and \cite{mashayek:2017_kh}).
Nevertheless, it is clearly necessary to continue
investigating whether and how the Thorpe scale can be related
to scales (and processes) of dynamical significance.

Just as it can be argued that the Osborn-Cox model is more inherently
appealing as a model for mixing since it relies
exclusively on properties of the structure of the density distribution, so too can an
argument be presented that $L_O$ is not
the most appropriate length scale to describe
 mixing, as it is
determined
by properties of the fluctuating velocity field rather than properties
of the fluctuating density field. The natural analogous length scale  is the
so-called
``Ellison scale'' $L_E=\rho'_{rms} /\left | \partial \overline{\rho}/\partial
        z \right | $
where $\rho'_{rms}$ is
the rms value of the density fluctuation away from $\overline{\rho}$,
(naturally closely related to the density variance associated with the
definition of $\chi$)
and
it is assumed that an appropriate characteristic
value can be identified from the spatio-temporally varying density
distribution.

Operationally, and similarly to the above-mentioned
Thorpe scale, the Ellison scale is straightforward to calculate from
a time series of measurements at a fixed location. As discussed by \cite{ivey:2018}, at least for energetic flows where the turbulence
is strong relative to the stabilising effects of 
stratification, there is
strong
observational evidence that $L_E$ is correlated
to a characteristic ``mixing length'' of stratified flows, and thus
$L_E$ proves to be potentially very useful as a length scale to describe
mixing. Nevertheless,  further investigation is undoubtedly needed to cement
the relationship between $L_E$ and nondimensional
parameters necessary for the construction of appropriate parameterizations.
This is yet another example of
an open, yet important question in the fascinating and environmentally
relevant research area of turbulence and ensuing mixing in stratified
flows.

\subsection{\modif{Outlook}}

\modif{ While there continues to be considerable advances in the research understanding of turbulence in stratified flows, using DNS and field observations particularly,
there is a growing gap between these advances and their implementation into predictive and managment tools developed to address UN Sustainable Development  Goals. For example, 
  these advances have not yet been appropriately incorporated into large scale ocean circulation models, particularly those running at global scales and on climate-change timescales. For example, in their global ocean model \citep{holmes:2019}, Holmes {\it et al.} (2019) parameterize diapycnal mixing using the
  deeply-influential``KPP'' model~\citep{large:1994} suggested  by Large {\it et al.} (1994) more than 25 years ago. This model assumes that the diapycnal diffusivity is simply a function of $\overline{Ri}$ - an attractive assumption for models with restricted vertical resolution and with heavy computational demands due to the model scale and time duration. But, as discussed above $\overline{Ri}$ is principally significant for determining the stability of parallel shear flows, not as a measure of the intensity of the mixing that may occur after the flow goes unstable~\cite{zaron:2009}. Furthermore, recent fluid dynamical research suggests that the concepts of ``marginal stability'' and ``self-organised criticality'' are significant, implying that flows often tune towards $\overline{Ri} \sim 1/4$, thus reducing the usefulness of
  a parameterization based around this parameter.}

\modif{There are ongoing controversies in the description of stratified mixing, even in highly idealized flows,
\modif{and this}
highlights the grand challenge of transforming recent advances in fluid dynamics research into relatively simple but physically realistic parameterizations.
\modif{ Achieving this grand challenge will enable
large-scale models to} produce reliable predictions of future climate change (SDG $\#$13), and aquatic ecosystem models can become powerful tools for ensuring clean water supply, sustainable cities and healthy aquatic ecological systems (SDGs $\#$6, 11 and 14).}

\section{Ocean transport and pollution}
\label{LagrangianCoherentStructure}
\subsection{Introduction}

Climate challenges \modif{require} a deeper understanding of the human impact on the earth system. For example the chemical compounds introduced \modif{into the atmosphere and in the sea \cite{VIATTE2020} 
 have a huge impact. These contaminants  interact with the biological components of terrestrial and marine ecosystems in a complex way, and their persistence, fate and transport in the air and marine waters need careful analysis.  Environmental fluid mechanics has traditionally focused on the basic and applied studies related to natural fluid systems as agents for the transport and dispersion of environmental contamination. From a climate challenge perspective, these studies are fundamental in establishing the scientific basis for adaptation and mitigation actions/plans. Here we concentrate on two aspects of environmental fluid mechanics.} 

The first is connected to the understanding that transport occurs in coherent structures since the ocean is dominated by large scale turbulence, manifested by pervasive eddies that can transport substances \modif{over large distances, thus} remaining coherent for very long \modif{periods. Having knowledge in advance of the coherence time of ocean eddies might thus  reveal the substance transport pathways. The recent work by Brach {\it et al.} \cite{BRACH2018} is of particular importance, showing that anticyclonic eddies increase the accumulation rates of microplastics in the North Atlantic subtropical gyre, a well-known area of plastic accumulation \cite{LEBRETON2012}.}

\begin{figure}[htbp]
\centering
\includegraphics[clip,width=0.56\textwidth]{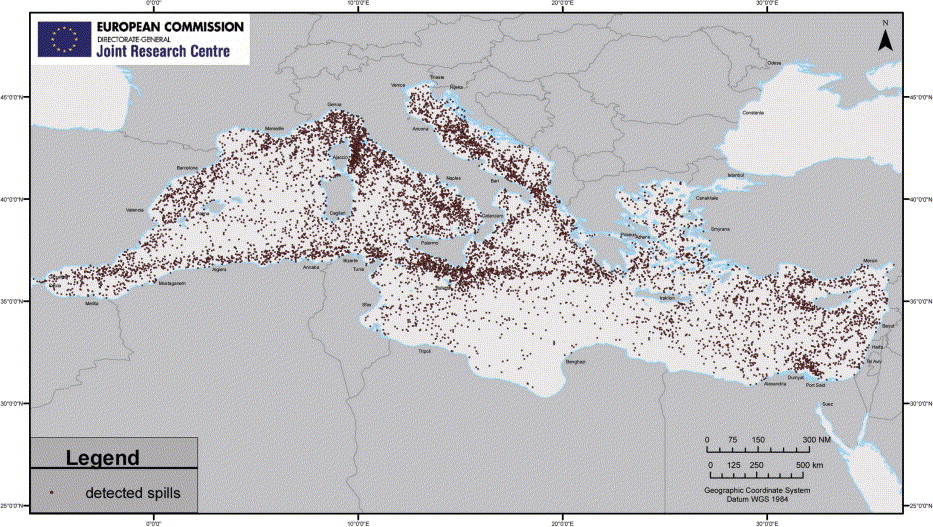}
\caption{\modif{Potential oil spills detected by satellite ESA Synthetic Aperture Radar (SAR) in the period 1999-2004 in the Mediterranean Sea \cite{FERRARO2007}.}} 
\label{fig:oilpollution}
\end{figure}

\modif{The second aspect} is related to the recent findings on the statistical distribution of oil pollution in the open ocean and coastal areas \modif{(SDG \# 14.1)}, which enable us to define a typical probability function for \modif{pollution transport and its arrival at the coasts.  Oil pollution at sea has the second highest contamination impact on the ocean due to the magnitude of maritime shipping. The volume of oil lost at sea from accidents amounts to 5.86 million tonnes~\cite{{ITOF2019}}, most of which is lost within 10 nautical miles from the shore. Although tanker spills have decreased by 90\% since the 1970s, they still occur and threaten the quality of the marine environment. Figure~\ref{fig:oilpollution} illustrates the estimated oil contamination in the Mediterranean Sea, giving an idea of the wide number of oil pollution sources within a 6-year period. This oil is transported by the turbulent oceanic flow field. The fate of the oil is related to the specific flow regime present at the moment the oil is released and the several days after. In studying oil dispersal at sea in a turbulent oceanic flow field, it is fundamental to understand  the probability distribution of the oil at sea and its arrival at the coast. The emerging statistical distributions for oil contamination at sea  enable appropriate indicators to be developed for monitoring and assessing acceptable limits of ocean pollution.}

\subsection{{Grand} Challenges for tracer transport structures}

\subsubsection{{The present status}}
Studies of ocean transport generally focus on nowcasting or forecasting
the evolution of scalar fields carried by currents. More often than not,
the objective of \modif{such} studies is not a highly accurate, pointwise prediction
of these scalar fields, but rather an identification of major pathways
to scalar field transport. Such pathways are most efficiently characterized 
by their boundaries, {\it i.e.} by transport barriers.

Geometric templates formed by transport barriers, such as fronts,
jets and eddy boundaries, are \modif{in deed} routinely used in geophysics
to describe flow features \cite{weiss08}. These templates are generally
inferred from instantaneous Eulerian quantities, even if the original objective
 is to characterize Lagrangian ({\it i.e.} material) transport.
This is often unsatisfactory because in turbulent flows, such as the
ocean and atmosphere, instantaneous Eulerian templates ({\it i.e.} velocity-field
based)  can yield transport estimates that differ by orders
of magnitude from actual material transport~\cite{haller13}.

The reason for this vast mismatch is twofold. First, material transport
is affected by the integrated effects of unsteadiness and trend changes
of trajectories in a turbulent flow. As a consequence, instantaneous
information from the velocity field and its derivatives does not account
for material transport over an extended time period. Second, \modif{according to} one
of the main axioms of continuum mechanics, descriptions of material
responses, including material transport, of any moving continuum should
be observer-indifferent
~\cite{Gurtin10}.
\modif{However,} the Eulerian diagnostics typically used in oceanography \textendash streamlines,
the norm of the velocity or vorticity and the Okubo-Weiss parameter
\cite{okubo70,weiss91}\textendash~ are all dependent on the observer.
This is at odds with a long-standing view in fluid mechanics that
flow-feature identification should be observer-independent \cite{drouot76a,drouot76b,astarita79,lugt79,haller05}. 

These discrepancies suggest that a self-consistent analysis of scalar
transport in the ocean should be carried out with objective Lagrangian
tools. Such tools \modif{could} be based on the mathematical analysis of partial
differential equations (PDE) of the advection-diffusion type, \modif{but this approach would be}
 hindered by the complex spatio-temporal structure of the velocity
field responsible for the advective component. \modif{One} alternative could be the
numerical analysis of the advection-diffusion equation, \modif{but that would be} similarly
challenging due to large concentration gradients near barriers and
generally unknown initial and boundary conditions. 

All these challenges often prompt transport studies to neglect diffusion
and consider only the advective transport of matter and properties.
In the absence of diffusive transport, however, transport barriers 
become ill-defined, given that any material surface completely blocks purely
advective material transport \cite{haller15}. This ambiguity
has resulted in the development of several \modif{alternative} theories for purely
advective transport barriers (Lagrangian Coherent Structures or LCSs),
with most of these methods identifying different LCSs even in simple
flows \cite{hadjighasem17}. 

\begin{figure}[h]
\centering 
\includegraphics[clip,width=0.56\textwidth]{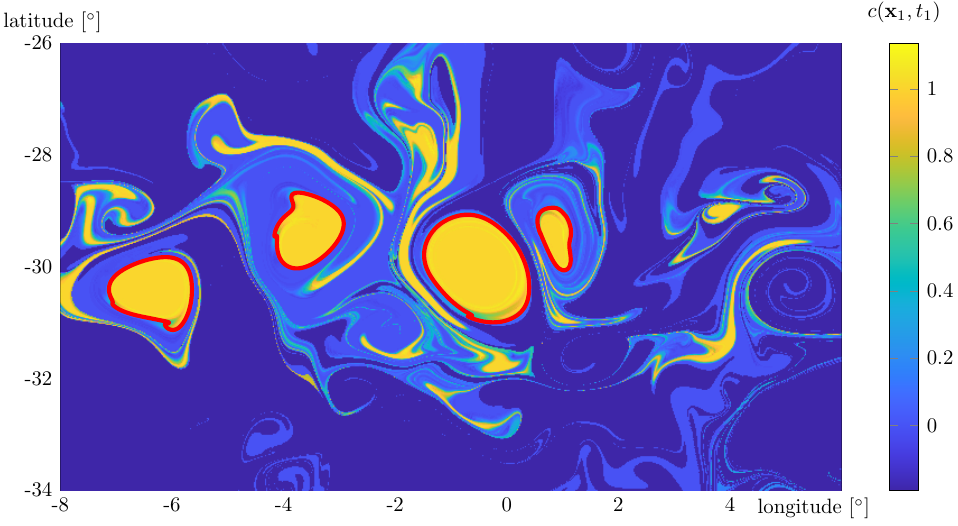}
\caption{Diffused concentration $c(\mathbf{x}_{1},t_{1})$ at time $t_{1}=t_{0}+90$ days,
with the advected positions of material Agulhas ring boundaries (identified
as diffusion barriers) overlaid \modif{in red}. Adapted from~\cite{haller18}.}
\label{fig:ocean-adv-diff} 
\end{figure}

As an alternative to LCS-based advective transport analysis,
one may seek \modif{transport barriers in turbulent flows} as exceptional material surfaces that block diffusive
transport more \modif{effectively} than any neighboring material surface \cite{haller18,haller19}.
\modif{Diffusion} barriers defined
in this fashion are independent of the observer
~\cite{haller18}. These results also
extend to mass-conserving compressible flows \cite{haller19} and
to barriers to the transport of probability densities for
particle motion in an uncertain velocity field modeled by an Itô process.
Figure \ref{fig:ocean-adv-diff} shows the application of these results to the extraction of
closed material barriers to diffusion that surround Agulhas rings
in the Southern Ocean. The algorithm \modif{that implements} these results for arbitrary two-dimensional
flows is available in BarrierTool, an open source MATLAB GUI downloadable
from \texttt{github.com/LCSETH.}

\subsubsection{Perspectives \modif{on} barrier detection}

These 
 results show the power of advanced variational calculus
to reconstruct key elements of a material transport-barrier network
from well-resolved numerical and experimental velocity fields. 
\modif{The barriers obtained in this fashion turn out to be coherent, but their construction is independent of any particular notion of advective coherence.}
\modif{They are constructed instead} from the universally accepted quantitative notion
of \modif{diffusive} transport through a surface. In the limit of the pure advection
of a conservative tracer, the theory renders material barrier surfaces
that will emerge as diffusion barriers under the addition of any small
diffusivity to the scalar \modif{field} or the slightest uncertainty to the velocity
field. 

Further challenges to address in this approach include an efficient
computational algorithm for transport barrier surfaces 
in three-dimensional flows, as well the inclusion of reaction terms
and coupling to other scalar fields. 
In addition, approximate versions of the exact theory of diffusion barriers
should be developed for sparse, observational data. A first step is the use of the diffusion-barrier strength 
diagnostic \cite{haller18}, a simple tool to locate barriers present
in the flow without computing null surfaces \modif{stipulated by the full theory}. 
Further steps might benefit from the use of machine learning
in the construction of barriers \modif{extracted} from under-resolved data, relying
on training a barrier detection scheme on highly-resolved data.

A further open question is the definition and detection of barriers
to the transport of active scalars, such as vorticity, potential vorticity,
helicity, linear momentum and energy. While the transport of these
active scalar fields is considered fundamental for building
the correct physical intuition regarding the flow, active scalars,
and measures of their transport, are 
observer-dependent, and hence their connection
with material transport is a priori unclear. A possible first step
would be to redefine these quantities so that they become objective,
or isolate a unique component in their transport that is observer-independent. 
\modif{
This approach has very recently been applied to the vorticity and to the linear momentum~\cite{HALLER2020}, but remains to be carried out for the helicity and the energy. }

\subsection{{Grand Challenges for \modif{oil pollution in the ocean}}}

\subsubsection{Distribution of ocean \modif{contaminants}}

Ocean \modif{contaminants} are distributed unevenly throughout the oceans and, as shown in the previous section, \modif{can be trapped or released} 
by eddies at different temporal and spatial scales \cite{eddies, Pearson-Baylor}. This intermittency of the oceanic flow field \modif{fundamentally affects} passive 
and active tracer transport , as described first by \cite{pierrehumbert94}. In \modif{his} seminal paper, Pierrehumbert described the probability density function (PDF) of passive and active tracer concentrations and found that they have exponential tails, {\it i.e.} they admit a tail with very large concentrations that depends on the specific turbulent flow field characteristics. 

If we apply this statistical analysis to ocean pollutant distributions, we\modif{ can objectively intercompare} the transport of tracers across basins with different current regimes, mean currents, mesoscale and submesoscale features, including the continental shelves of the world’s ocean basins, where the dynamics are different from the open sea. Ultimately the statistical representation of pollutant advection-diffusion transport in the ocean \modif{guides us in formulating general
indicators for climate change challenges related to environmental contamination.}

\modif{Oceanic and atmospheric dynamical fields, as well as the environmental tracers dispersed in the atmosphere and at sea, show PDFs that} are normally represented by two parameter distributions \cite{Prashant2015}. 
different tracer advection-diffusion regimes \modif{can be reduced} by describing how these parameters vary in different regions and at different times.  PDFs for world-ocean currents have been calculated from satellite altimetry \cite{Chu} and numerical circulation models~\cite{Ashkenazy}.  For tracers, \cite{Hu} assessed the advection-diffusion PDFs for stratospheric tracers and \cite{Sepp-Neves} \modif{did the same} for oil in the ocean, both papers using realistic numerical simulations.  The emerging PDF for both currents and pollutants is of Weibull type, {\it i.e.} it can be written as 
$P(x;\alpha,\beta)= ({\alpha}/{\beta})\left({x}/{\beta}\right)^{\alpha-1}\exp[{-\left(\displaystyle {x}/{\beta}\right)^{\alpha}}],$
 where $x$ is the tracer concentration, $\alpha$ is the shape and $\beta$ is the scale parameter. \modif{The PDF parameter values will most likely} vary slowly in time. This PDF is characterized by a Gaussian core and fat tails, which fall more slowly than a Gaussian, \modif{and anomalously indicate the} high probability of extreme concentration fluctuations. This means that mixing or diffusion do not act fast enough to homogenize the tracer, which remains at a high concentration for a finite-time.

\modif{Ocean ensemble simulation} approaches are effective to study the \modif{PDFs of pollutants} because monitoring of ocean tracers is still difficult both from satellites and in situ. This is in contrast with the atmosphere in which most tracers can be observed from space. In particular, for plastics \cite{MAXIMENKO,LIUBARTSEVA} and accidental and operational oil releases \cite{LIUBARTSEVA2015, SeppNeves2016}, simulation-ensemble techniques are emerging methods to study hazards from pollution. Ocean-ensemble simulations currently benefit from the best reconstructions of ocean currents from operational ocean forecasting centres, which provide multi-decadal time series of the ocean flow field~\cite{LETRAON2019}. This ensemble-statistical framework is also very important in accounting for uncertainties in the tracer release positions, errors in current reconstructions, and errors in the chemical and physical transformations represented in active tracer dynamics. 

\begin{figure}[h]
\centering
\includegraphics[width=2.8in]{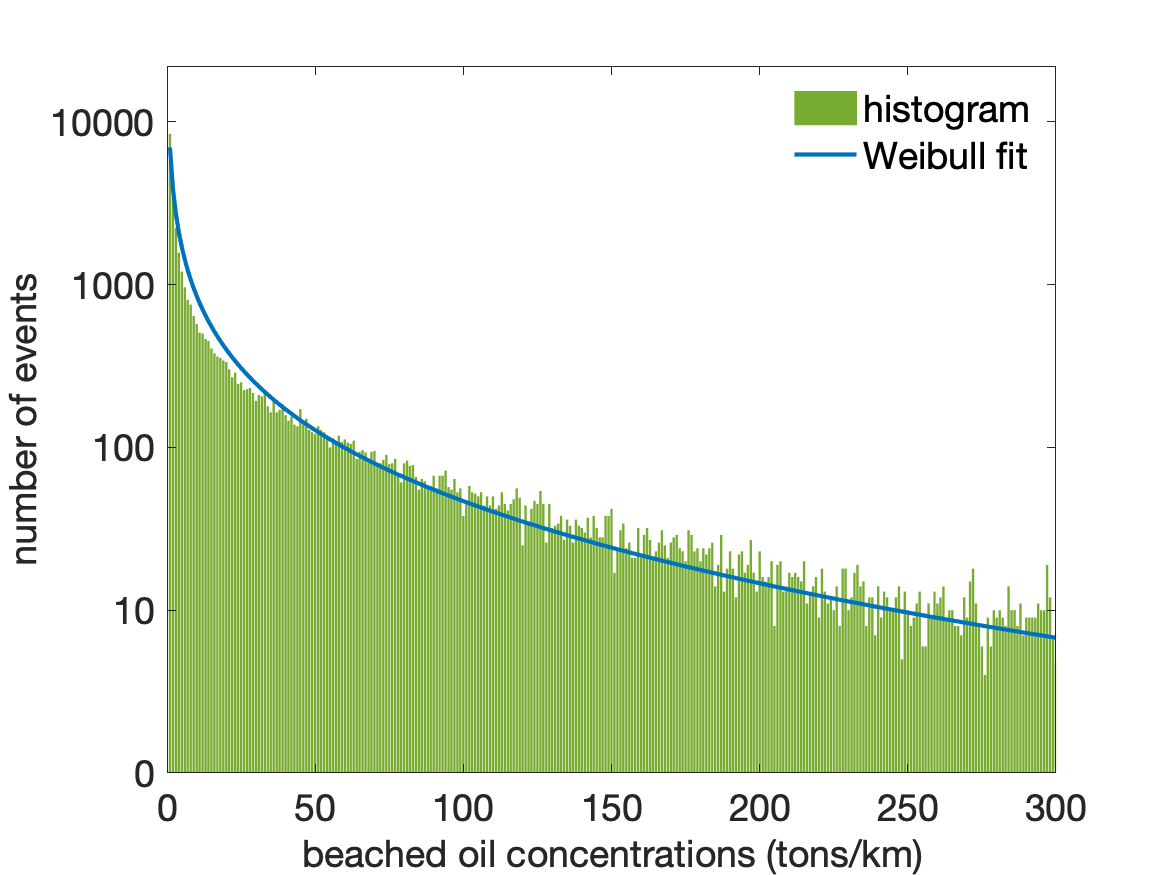}
\caption{Beached oil distribution from ensemble simulations for different release points around the Caribbean Islands, simulating accidental oil spill transport produced by the flow field conditions in 2013.}
\label{fig:oil-PDF}
\end{figure}

Figure~\ref{fig:oil-PDF} shows a distribution for beached oil concentrations for the entire Caribbean Archipelago coastline \modif{using an ensemble simulation approach with different virtual release points and one year of the realistic flow field from the Copernicus Marine Environment Monitoring and Forecasting Service~\cite{LETRAON2019}.} The long tail of high concentration values \modif{highlights} the importance of understanding PDF distributions \modif{in calculating hazards accurately. A high concentration of oil could arrive at the coasts even from single release points around the islands, depending on the flow field structures which depend on the dynamics of currents in the area for a given amount of time.}

\subsubsection{\modif{Applications of ocean oil pollution PDFs}}

Hazards from oil pollution \modif{stem} from the relatively high number of events in the distribution tail \modif{of the PDF just described}. The Weibull tail distribution, $H$, \modif{can be used to quantify the hazard since it is}  the integral of the PDF bounded by an appropriately chosen low-value concentration, $x_{low}$, as 
$H(x_{low};\alpha,\beta)= \exp[{-\left({x_{low}}/{\beta}\right)^{\alpha}}].$
\modif{This function  is used to map beached oil spill} hazards due to oil releases from maritime traffic or accidents. 
\modif{The ensemble simulation consisted in generating several hundred thousand simulations using different high resolution flow fields from oil release points in the sea area from land to 100 km offshore. 
Using a global approach to oil pollution hazard mapping, we explored the values of $H$ for different coastline segments across the whole North Atlantic region from these ensemble simulations. Table 1 presents the $H$ values for five different areas with a threshold of $x_{low}=25$ tons/km. The $H$ values are sufficiently different to characterize the different hazards of beached oil in the different coastal segments. This means that beached oil PDFs are useful to characterize hazards that might be transported toward the coasts due to the different current regimes. In this generalized view of hazard mapping from the study of ensemble simulated oil contamination, we should soon be able  assess high and lower hazard coastal segments in the global ocean.}

\modif{Preventing and significantly reducing} marine pollution of all kinds \modif{(SDG \# 14.1)}, including marine debris and nutrient pollution, \modif{could} be described by the \modif{PDF of these tracers. Thus, all sea contamination hazard mapping could   solely be based on the study of the relevant PDF and its parameters. Monitoring with PDF parameters or derived quantities, such as the $H$ index from ensemble simulations, could be used as the basic method for periodically assessing the degree of pollution} in the world oceans.

\begin{table}
\centering 
\begin{tabular}{| l | c  |}
\hline
Coastline segment & $H$ index values      \\
\hline
\hline
Atlantic French     & 0.8   \\
Madeira Island   & 0.33     \\
Bahia region (Brazil)     & 0.16  \\
Mexico   & 0.17 \\
US North Atlantic & 0.20 \\
\hline   
\end{tabular}
  \caption{\modif{The hazard index $H$ calculated from an ensemble based simulation of oil releases at sea for one year of realistic currents.}}
 \label{io }
\end{table}

\subsection{Outlook}

\modif{Many basic phenomena and processes in the transport and dispersal of ocean contaminants still need to be clarified and require future investigation. As outlined much more comprehensively by Ba{r}ker {\it et al.} \cite{Barker2020}, oil pollution science requires an improvement in oil model transformation, a better consideration of ocean currents and winds that affect the fate, transport and the development of new numerical methods for the representation of oil transport, i.e. Lagrangian particles versus bulk concentration models.}

\modif{Above all, a better presentation of transport using three-dimensional ocean currents is key: horizontal and vertical resolution should be increased to enable mesoscale and submesoscale dynamics to be resolved, including tidal currents and Langmuir vertical circulation and correctly accounting for turbulent mixing for these kinds of tracers.
The Weibull PDF, recently discovered for oil pollution in the ocean, is likely connected to the material transport barriers described in the previous section and to other characteristics of the oceanic and atmospheric turbulent flow field, which will be developed in future research.}

\modif{Another problem that requires further investigation is related to the appropriate sampling of uncertainties by ensemble-based simulations. For many contaminants, uncertainties are related to the unknown size and modalities of the contaminant release, the type of oil contaminant, the position(s) of the initial release and the variability of the wind and current flow fields. This represents a formidable challenge to the number of simulations required to sample the uncertainties in a comprehensive way and to manage the methods and analyze the model output data.}

\modif{Last but not least, machine learning from the vast data sets available from simulations and data intensive field expeditions may also lead to very significant progress in predictive capabilities and hazard mapping~\cite{GROSSI2020}.}

\section{Urban Flows}
\label{PredictingUrbanFlows}

\subsection{Introduction}
By 2050, 6.5 billion people, or two-thirds of humanity, will live in cities. This rapid urbanization brings enormous challenges, thereby motivating SDG $\#$11: to make cities and human settlements inclusive, safe, resilient and sustainable. To achieve this goal, significant transformations will be required in the way cities are designed, managed, and built~\cite{UN2019}. Urban fluid mechanics plays an important role in ensuring the safety, resiliency and sustainability of cities: the wind patterns in the urban canopy affect structural resiliency, pedestrian wind comfort and exposure to pollution, street canyon ventilation and air quality, wind energy resources, natural ventilation of buildings and indoor air quality, and urban heat island effects. \modif{The negative economic, environmental and equity consequences of poorly managed urban wind effects are enormous. For example; the US recorded a \$24 billion insured loss due to extreme wind events in 2019 ~\cite{III2020}; without action to address energy efficiency, energy consumption for space cooling is projected to more than triple by 2050, consuming as much electricity as China and India today
 ~\cite{IEA2018}, and; communities with low socioeconomic status experience higher concentrations of air pollutants, resulting in higher respiratory and cardiovascular disease rates ~\cite{Hajat2015}.}

\modif{Urban flows can include multi-phase flows and scalar transport, as well as stable and unstable stratification. In addition, urban flow has a fundamentally multi-scale nature, governed by large-scale weather patterns down to Kolmogorov microscale turbulence. As such, one can draw many parallels between the grand challenges described in Sections~\ref{Multiphaseflow},~\ref{Stratifiedflows},~\ref{LagrangianCoherentStructure} and~\ref{Prediction} and the challenges faced in improving our fundamental understanding of different urban flow problems. Instead of elaborating on some of these challenges in the context of urban flows, this section will focus on the overall grand challenge of predicting urban canopy flows. This focus is motivated by the vision that accurate urban flow predictions could support the design and engineering of urban areas and buildings to not only mitigate negative effects or adapt to the consequences of climate change, but to actively create an environment that equitably improves city dwellers' lives.} In the following we first outline the grand challenges towards enabling accurate predictions, before summarizing recent progress on case studies considering natural ventilation and urban flow and dispersion.

\subsection{Grand Challenges in Predicting Urban Flow}
\label{subsec:urbanchallenges}

\modif{Physical experiments and computational models each have an important role to play in improving our understanding of urban flow, but the complexity of urban flows limits their individual predictive capabilities. Specifically, three grand challenges can be identified: representing the complexity and heterogeneity of urban geometries, accounting for the inherent variability in urban flows, and accounting for uncertainty in reduced-order physics models in computational tools.} This section aims to summarize the effect of these challenges on the predictive capability of laboratory measurements and computational models, thereby identifying the need for novel approaches that integrate both methods with field measurements, which represent the full complexity of urban flows.

\subsubsection{Representing the complexity and heterogeneity of urban geometries}

Urban flow is governed by a wide range of scales: the wake downstream of a city downtown area can be a few kilometers, while the smallest scale, determined by the Kolmogorov microscale, is on the order of millimeters. In between, there is a range of geometrical features, such as the overall building dimensions and spacing, balconies and windows on building fa\c{c}ades, and vegetation, that locally influence the flow. Geometry-specific simulations or experiments aim to reproduce these effects, but the level of geometrical detail that should be represented remains an open question. It is well established that the aerodynamic effects of vegetation influence the urban wind environment~\cite{mochida2008}, and geometrical details in the urban canopy have been found to modify the local flow field~\cite{montazeri2013,llaguno2017}. The observed effects are often specific to the configurations and quantities of interest considered, indicating a need to develop generalized, systematic approaches to define the required accuracy and level of detail in the geometrical description. Such approaches should weigh the potential improvement in the accuracy of the predictions, which comes at an increased computational cost, against the uncertainties introduced by the other two challenges.

\subsubsection{Accounting for inherent variability in the boundary and operating conditions}
Urban flow studies have traditionally employed carefully scaled laboratory experiments in atmospheric boundary layer (ABL) wind tunnels. These wind tunnel tests are routinely used to inform building design and validate computational fluid dynamics (CFD) simulations, even though it is recognized that there is a lack of validation with full-scale field measurement data~\cite{baker2007}. Several studies comparing wind tunnel and field experiments have identified non-negligible differences between measured quantities of interest, including the wind speed and direction, the concentration of pollutants, and the wind pressure on building fa\c{c}ades~\cite{klein2007,schatzmann2011,asghari2016}. The inherent variability in the real ABL has been cited as an important reason for these discrepancies: the boundary conditions of a field experiment cannot be controlled, and larger-scale variability in the ABL prohibits the acquisition of time-series representative of the quasi steady-state flow conditions in the wind tunnel. When modeling flow in buildings, additional uncertainties arise due to continuous changes in operating conditions, such as occupancy and the corresponding heat loads that determine buoyancy-driven flows. 

To improve our understanding of the effects of this inherent variability and validate predictions with full-scale data, there is a need for novel probabilistic modeling strategies and for detailed field measurements. Deterministic, point-wise, comparisons have been inconclusive due to the limited amount of data that can be obtained for both the quantities of interest and the characterization of the boundary and operating conditions during the experiment. Probabilistic approaches that can represent the effect of the variability in the field have been shown to provide a more meaningful comparison~\cite{harms2011,asghari2016}, but can be time-consuming in the lab; instead, advances in high-performance computing capabilities, numerical algorithms, and tools for uncertainty quantification, can enable efficient evaluation of the effect of the inherent variability in computational models. Important research questions regarding the definition of probability distributions for the uncertain parameters and the most efficient way to propagate them to the quantities of interest remain. The answers to these questions will be different for different quantities of interest, and carefully designed field experiments are required to further develop and validate probabilistic approaches. These experiments should not only gather data for relevant quantities of interest, but also characterize probability distributions of variable boundary and operating conditions that could affect these quantities of interest.

To further improve the realism of ABL inflow boundary conditions in CFD simulations, they can also be coupled to larger-scale weather forecasting models. The coupling of these codes is not straightforward; their different physics modeling approaches and the large disparity in the resolution of the simulations imply that some form of interpolation, model blending, or generation of smaller-scale turbulence is required~\cite{mochida2011,yamada2011}. The downscaling of weather forecasting codes to enable obstacle resolving simulations can alleviate the need for model blending, but the use of nested grids and immersed boundary techniques still has numerical and physical modeling challenges~\cite{chow2018}. Importantly, in both the coupled and downscaled simulation approaches, the quality of the solution will strongly depend on the accuracy of the larger-scale simulation~\cite{wyszogrodzki2012,talbot2012}. The grand challenges in weather prediction models are discussed in section~\ref{Prediction}; for the purpose of using their output to define boundary conditions for urban-scale CFD models, it will be essential to define strategies that propagate the uncertainty in the weather model prediction through the urban-scale model~\cite{garcia2018}.

\subsubsection{Accounting for uncertainty in reduced-order physics models}
The use of reduced-order physics models in numerical simulations introduces an additional challenge. For example, urban flow simulations generally employ some form of turbulence modeling to represent the effect of the large range of turbulence scales on the mean flow and on the transport of pollutants or heat. The choice of the turbulence model is essentially a trade-off between fidelity and computational cost: Reynolds Averaged Navier-Stokes (RANS) simulations offer a low-fidelity, affordable option, while large-eddy simulations (LES) provide a high-fidelity, expensive solution. Similar to the challenges encountered in the parameterization of mixing in stratified flows (section ~\ref{Stratifiedflows}), traditional comparison and calibration of RANS turbulence models with wind tunnel experiments for urban flows has only been moderately successful. As a result, the converging opinion is that we need LES for improved accuracy~\cite{blocken2014}. 

When considering validation with field measurements, this conclusion becomes more ambiguous. In wind tunnel validation studies, geometrical differences and variability in the flow conditions can be largely eliminated, such that turbulence modeling becomes the main challenge. In field measurements the other challenges can dominate, and the use of an expensive turbulence model no longer guarantees an accurate prediction~\cite{neophytou2011,garcia2018}. To ensure full-scale predictive capabilities, geometrical uncertainties and variability in flow conditions should also be represented. To achieve this within the limits of acceptable computational cost, one can not solely rely on LES. Instead, we need to explore new multi-fidelity simulation approaches, where an expensive high-fidelity model or experiment can be used to calibrate a fast low-fidelity model, and the low-fidelity model can then be used to quantify the effect of variability in the real full-scale conditions~\cite{peherstorfer2018}. In this context, research on the use of machine learning to quantify and reduce uncertainty in RANS turbulence models based on high-fidelity simulation data bases also has the potential to improve urban flow simulations~\cite{duraisamy2019}.

\subsection{\thie{Two example} Grand Challenges}
\label{sec:level2}
\begin{figure}[htbp]
\centering
\includegraphics[width = 0.55\textwidth]{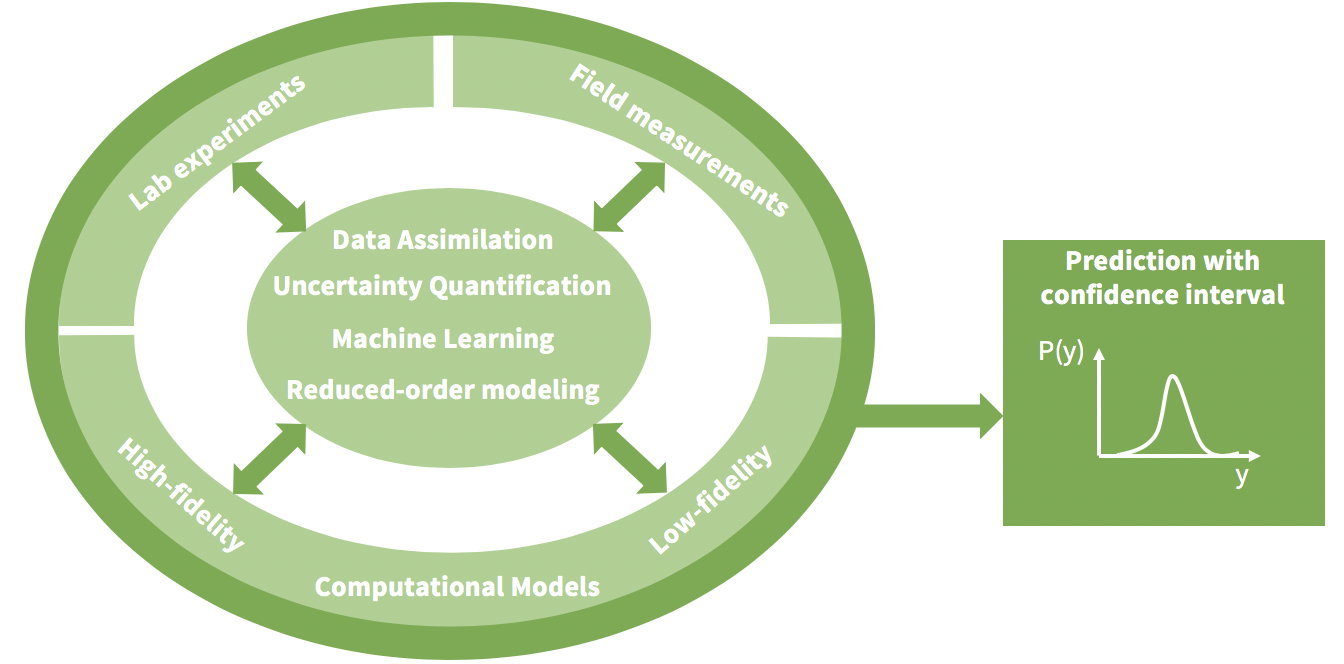}
\caption{Integration of field measurements, laboratory measurements, and high- and low-fidelity computational models to provide predictions with confidence intervals.}
\label{fig:framework}
\end{figure}

Figure~\ref{fig:framework} visualizes the different sources of data and the integration methods that can contribute to addressing the challenges identified in section~\ref{subsec:urbanchallenges}. This section presents recent progress on two different applications to illustrate how a subset of these techniques can contribute to improving our understanding of urban flows and achieving validation with realistic field data. Both studies demonstrate the need for field experiments and high-fidelity modeling, while also highlighting opportunities to use the resulting data to develop faster low-fidelity models that can provide predictions with confidence intervals to inform design.

\subsubsection{Natural ventilation}
\label{sec:flowanddispersion}

A major challenge posed by increasing urbanization is the huge and increasing energy demands of the built environment and the consequent greenhouse gas emissions and heat island impacts. Much of this energy use stems from the increasing use of air conditioning; the 2017 International Energy Agency report `The future of cooling' highlights concerns about an unsustainable energy demand for cooling associated with urbanization - the so-called ``cooling crunch''. An alternative approach is needed if urbanization is to be sustainable, and one possibility is to replace air conditioning with natural ventilation which uses the \emph{energy-free} resources of the wind and temperature differences between indoors and outdoors to drive ventilation flows through a building. {This is the objective of the Managing Air for Green Inner Cities (MAGIC) project (www.magic-air.uk)} \citep{song_etal18}.

In order to use natural ventilation it is necessary that the external air has an acceptable level of air quality both in terms of pollutants, gaseous and particulates, and is appropriate in terms of temperature and humidity. It is also necessary to have information on the external environmental conditions and the wind flow in order to ventilate buildings effectively and to provide comfortable conditions inside the building. In terms of day-to-day operation this may be achieved by having access to local monitoring data. On the other hand, in order to design naturally ventilated buildings, or retrofit existing buildings, and to place them in an urban context, requires a sophisticated modelling framework that provides a systems approach to this highly interconnected and complex problem. Such an approach must also account for variations in weather, traffic and other time-dependent patterns such as solar radiation, spatial variations in pollutant concentrations and occupant behaviour.

To achieve this, MAGIC employs field studies, laboratory experiments (wind tunnel and water flume on flow around and inside buildings), and high-fidelity modeling. Field studies carried out in London in 2017 \modif{and 2019}, and Cambridge in 2018 show that both indoor and outdoor pollutant levels are highly variable. The measurements clearly demonstrate the need for high-fidelity modelling. To this end, MAGIC employs the LES open-source code Fluidity which has an adaptive unstructured mesh that allows the highly localized computations of wind speed, temperature and pollutant levels needed to evaluate the performance of a naturally ventilated building within its particular urban context. Fluidity allows for neutral, unstable and stable ABL flows, and includes thermal radiation from buildings and sensible and latent heat transfers from green and blue space, an urban design term that stands for visible water. While Fluidity has the capability to make the required calculations it is computationally expensive and has long run times. Consequently, MAGIC also employs data assimilation, reduced order modeling and machine learning to improve accuracy and to speed up run times so that calculations can be run in close to real time. The coupling of these technologies still represents a significant challenge but the present outlook is encouraging. For example, reduced order modeling produces speed-up by factors of 10$^6$, allowing for calculations to be used in design studies.

\subsubsection{Urban flow and dispersion}
\label{sec:natvent}
In 2016, 91\% of the world population was living in places where the world health organization air quality guidelines were not met, and outdoor pollution was estimated to cause 4.2 million premature deaths worldwide~\cite{WHO2018}. Detailed predictions of wind and dispersion patterns in urban areas could provide essential information to mitigate adverse health effects. However, the predictive capability of numerical models is limited by the challenges identified in section~\ref{subsec:urbanchallenges}. This case study attempts to address these challenges by (1) investigating methods to quantify the effect of uncertainty in the inflow boundary conditions, and (2) evaluating the relative importance of inflow and turbulence model uncertainties. Two different configurations were considered: the Joint Urban 2003 (JU2003) experiment in Oklahoma City~\cite{OklahomaCity} and a recent field measurement on Stanford's campus. 
\begin{figure}[htbp]
\centering
\includegraphics[width = 0.77\textwidth]{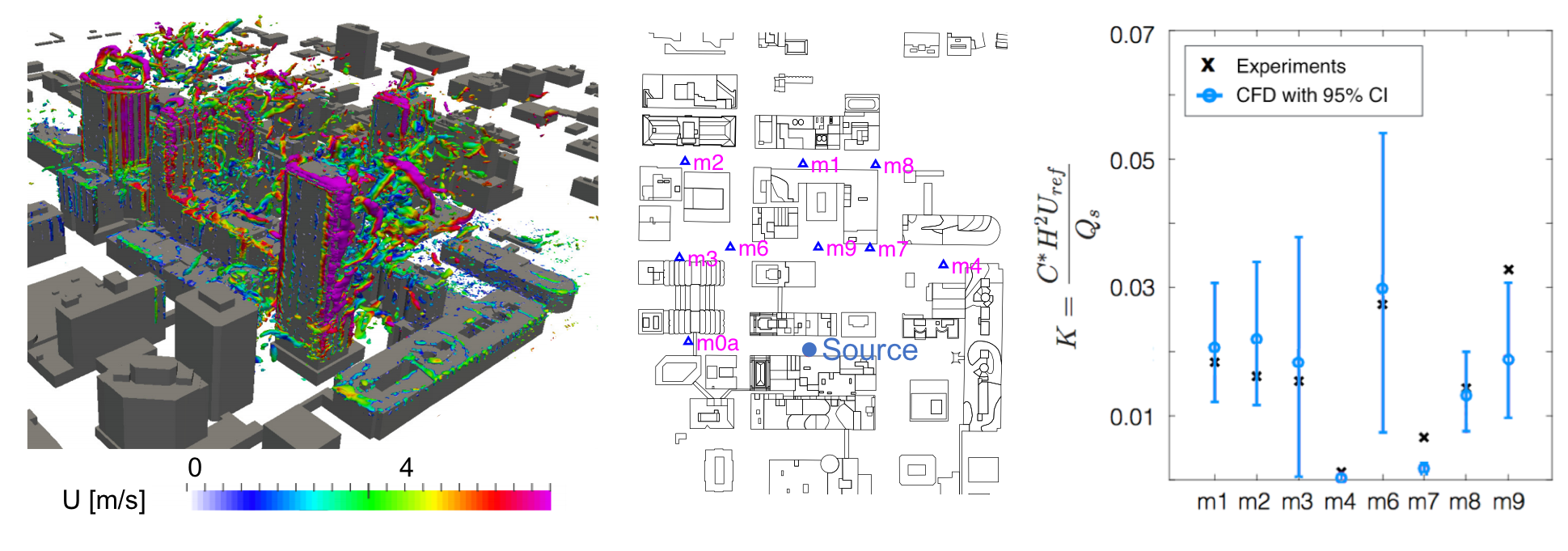}
\caption{Simulation results for JU2003: iso-contours of Q-criteria colored by velocity magnitude obtained from LES~\cite{garcia2018} (left), pollutant source and sensor locations (center), and RANS predictions of pollutant concentrations with 95\% confidence intervals compared to field measurements~\cite{garcia2017}. } 
\label{fig:OKCresults}
\end{figure}

To quantify the effect of the inflow uncertainty on the simulation results, three uncertain parameters were defined: the ABL roughness height, and the wind magnitude and direction. For JU2003, probability distributions for these parameters were defined using either field data from a sensor placed close to the inflow boundary~\cite{garcia2017}, or output from mesoscale simulations~\cite{garcia2018b}. The uncertainties were propagated to the quantities of interest using a polynomial chaos expansion approach. The results, shown in Fig.~\ref{fig:OKCresults} (right) for the study using the field measurements as input, indicate the potential of this approach when addressing comparisons with field measurements. The use of weather forecasting models to define the input distributions also provided realistic results, but the uncertainty in the predicted concentrations is significantly larger due to uncertainty in the mesoscale model output. This motivated an experiment on Stanford's campus to determine if using data from sensors inside the urban canopy could also provide improved predictions. Wind velocity data from two sensors inside the urban canopy were assimilated using an ensemble Kalman filter; data from four additional sensors were used for validation. The predicted mean values were~$\sim$20\% more likely to be within the 95\% confidence interval of the experimental data compared to the traditional method of using weather station data to define the inflow boundary conditions~\cite{sousa2019}. 

The relative importance of turbulence model form uncertainties compared to inflow uncertainties was investigated in two ways. First, a high-fidelity LES (Fig.~\ref{fig:OKCresults} (left)) was performed for the dominant wind direction during JU2003~\cite{garcia2018}. Comparison of the modeled and measured wind velocities indicated there was no tangible improvement in the LES predictions compared to RANS, indicating that the influence of other uncertainties can not be neglected. Second, an approach to quantify RANS turbulence model uncertainties by introducing perturbations in the modeled Reynolds stress tensor was explored~\cite{gorle2015}. The approach predicts a plausible interval for the quantities of interest; the magnitude of these intervals varied locally, but they were generally smaller than the confidence intervals predicted by the inflow uncertainty quantification study. Multi-fidelity approaches could offer further opportunities for decreasing the magnitude of the intervals: data from high-fidelity simulations or experiments could inform the perturbations introduced in the Reynolds stress tensor. 

\subsection{Outlook}

\modif{The case studies presented in this section highlight the complexity and high variability of urban flow problems. They clearly demonstrate the importance of accounting for the different types of uncertainties and the need for field experiments and high-fidelity modeling. The methods used in these studies represent a first attempt towards improving the predictive capabilities of the simulations; there are many open questions and opportunities for future research. The most exciting opportunities might lie in the increasing amount of data that can be obtained from urban sensor networks and from high-fidelity simulations, and in exploring new algorithms to integrate this data with low-fidelity models that are sufficiently fast and robust to inform design and policy decisions.}

\modif{Finally, there are significant challenges in translating urban flow research results into impacts on people. Stakeholder interaction at the urban dweller or building occupant, designer or engineer, and municipality or developer levels will be required. An important goal of these interactions should be to better understand and characterize the economic, equity, and environmental benefits of sustainable design solutions to inform effective policy. For example, personal exposure to pollution is highly variable, both outdoors but also indoors where we typically spend 90\% of our time. The impacts of this exposure and other aspects of the urban environment, such as the access to daylight, green spaces, and `fresh air' on human health, well being and productivity is not well understood, yet critical to living fulfilled lives in cities.} 

\section{Weather and Climate Prediction}
\label{Prediction}

\subsection{{Introduction}}

The prediction of the atmospheric state is key for all socio-economic sectors that depend on weather and air quality, and climate change adds significant complexity to the problem through anthropogenic contributions that are measurably affecting our planet. Despite the skill of today’s forecasting, tens of thousands lives and hundreds of billion dollars are lost due to weather extremes every year \cite{UNISDR2018}. This clearly asks for much enhanced predictive skill and an assessment of where opportunities and challenges lie.

\modif{This makes weather and climate prediction highly relevant for a number of {SDGs}, in particular {SDG $\#1, 2, 3, 6, 7, 13, 15$} as weather extremes and how such extremes will evolve under climate change affect all aspects of our water, energy and food supplies with huge implications on health, economy and environmental-stress induced social and political instability. Future policies on carbon-neutral societies like the European Green Deal rely on reliable information for decision making. Reliable weather and climate prediction capabilities form an integral part of such policies.}

Today’s most sophisticated prediction systems include atmosphere, oceans, sea-ice, land surface and key components of the biosphere since the Earth-system is a high-dimensional, non-linear dynamical system in which all of these components interact at different space and time scales. Predictive skill depends therefore on how realistic the Earth-system physics are represented in models, and how well this system can be observed to formulate the underlying physical laws, and how well accurate initial conditions and external forcings for forecasts can be derived.

Historically, weather and climate prediction have diverged because weather models focused on shorter time scales (days to months) while climate models on longer scales (decades to centuries, or even millennia for paleo-climate studies) \cite{Lynch2008}. Due to computing cost, this choice had implications on model resolution and complexity, so that climate models operate at best at $O($25 km) today but include all Earth-system components \cite{Haarsma2016}, while weather models operate at $O$(10 km) with much more physical process detail but an incomplete representation of the Earth system \cite{WGNE2018}. Another major difference is that weather models need very accurate initial conditions while climate models are only weakly initialized \cite{Doblas-Reyes2013,Taylor2012}. The weather application has also pioneered the concept of ensemble prediction, which adds a physically based uncertainty estimate to initial conditions and forecasts~\cite{Palmer2005}.

However, this historic separation is about to end because of the generic need for more realism in model physics, the essential role of observations in identifying model errors, and the technological limitations of high-performance computing and big data handling. All three present Grand Challenges for Earth system prediction are highly interconnected, and their solution will require non-traditional ways of thinking. 

\subsection{{Grand Challenges for model physics}}
Global prediction models are based on a set of equations describing three-dimensional motion, the continuity equation, and thermodynamic and gas laws. While these equations may be formulated around different prognostic variables and coordinate systems, they accurately represent the fluid flow. As the equations cannot be solved analytically, they require numerical methods to advance the state of prognostic variables in time and space. These methods rely on various grid set-ups, and have different implications on conservation, balance, and numerical stability and accuracy. This part of the model is usually called the ‘dynamical core’, and it describes the dynamics of processes that are resolved with the chosen discretization \cite{Cote2015}.

A unique aspect of weather and climate models is the need to parameterise the impact of sub grid-scale processes on mass, momentum and energy advanced at the resolved scale. In weather models, examples of such processes are radiation, convection and clouds, surface drag and gravity waves excited by orography and in the free atmosphere, and the interaction between the atmosphere and surfaces \cite{Brown2015}. The coupling to land and vegetation, ocean, wave, sea-ice and ice-sheet models is carried out by exchanging fluxes at the interfaces.

‘Parameterisation’ means that many of these processes are represented by approximate laws often derived from observations with limited representativeness. Prominent examples are deep convection, clouds and orographic drag – all being of very high importance for predictive skill. Maintaining approximate laws in physical models is considered a key impediment to progress \cite{Bony2015}, and hence eliminating parameterisations by actually resolving the full process is clearly an option for consideration.

While predictive skill of weather models has steadily increased over time \cite{Bauer2015}, and climate models show enhanced agreement with observations when run over past periods, adding complexity by including more and more physical and chemical detail has not led to the elimination of key skill limitations in recent decades \cite{Stevens2013,Bellprat2016}. Enhanced resolution has clearly shown benefits \cite{Roberts2018} but there is evidence that this improvement is not steady and that there are key resolution thresholds that need to be overcome to reliably predictive key Earth-system mechanisms \cite{Prodhomme2016}. Past examples are resolutions better than 100 km to resolve mid-latitude frontal structures \cite{Zappa2013}, 20-40 km that helped resolving the complex scale interaction in weather regime transitions, for example blocking \cite{Jung2012}, and at least 50 km for representing the inter-annual variability of tropical cyclones \cite{Roberts2015}. 

\begin{figure}[h]
\centering
\includegraphics[width=5.0in]{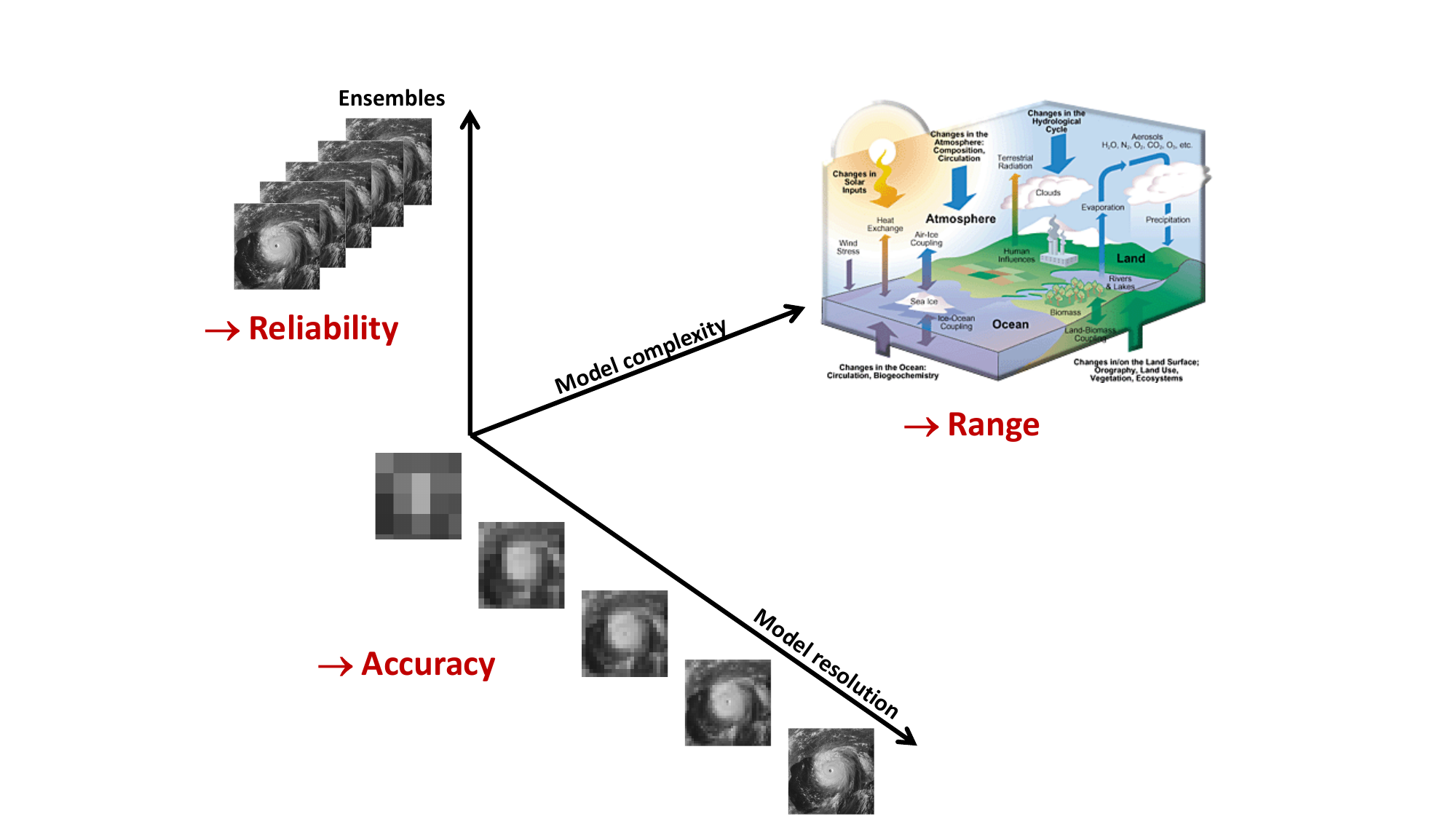}
\caption{{Future prediction system advances will arise from improved resolution delivering a more accurate representation of physical processes, more complex models delivering a better representation of the entire Earth-system that is highly relevant for longer-range prediction, and larger ensembles delivering more reliable forecasts from better uncertainty statistics. Increasing both ensembles and model complexity has at least a linear impact while better resolution has at least a cubic impact on the computational cost of simulations.}}
\label{fig:oil-PDFbb}
\end{figure}

However, shifting the boundary between resolved and parameterised processes by a significant step appears to be the only way to overcome key sources of model biases: this is the first big challenge. Numerical experiments with very high-resolution models indicate that deep convection in the tropics must be resolved to accurately describe convection dynamics and its effect on the large-scale circulation, which drives weather patterns at all latitudes \cite{Bony2015}. Shallow convection and stratified cloud processes in sub-tropical areas represent the next barrier as these clouds determine an important contribution to the global energy balance via radiation, and exhibit strong sensitivity to heating trends in the atmosphere following climate change \cite{Bony2005}. 

Surpassing both barriers implies running global models at 100 m - 1 km resolution, which seems to present a nearly impossible computing task \cite{Wehner2011,Schulthess2019}. Requiring such enhanced processing capability translates into a much closer co-development between model physics, numerical methods and their implementation on highly parallelised and energy efficient hardware. This is common to both weather and climate models.

\subsection{{Grand Challenges for observations}}
Traditionally, observations have been used for model and forecast verification and, through dedicated observational field campaigns and reference stations, also for model development \cite{Bauer2015}. The weather and climate community is very well organised in defining their observational requirements, common observational network strategies, supporting future satellite programmes, and exchanging data globally in near real time with unified formats and metadata. This effort is one of the key foci of the World Meteorological Organisation and space agencies, and is strongly supported by national and collaborative efforts across countries.

Today’s operational weather forecasting centres employ about 60 million observations per day  for generating initial conditions for forecasts and for verification based on data assimilation methods. Similar data volumes are being employed for climate and air-quality reanalyses supporting climate monitoring and predictions~\cite{Dee2014}. The accuracy of the initial conditions is largely determined by the quality of the forecast model as observational information can only be exploited when the forecast model produces a state estimate that is close to the observed one. The above model development challenge therefore projects directly onto data assimilation. At scales of 100 m - 1 km, so-far parameterised processes will be resolved so that also data assimilation methods need to be able to exploit high-resolution observations, represent small-scale and fast processes, and describe interactions across a wider range of time and space scales. 

While climate projections beyond decadal time scales are not initialised with observed data, there is significant potential to exploit data assimilation methods for model development serving both weather and climate prediction \cite{Phillips2004}. Firstly, systematic forecast errors appear very early in the forecast so that error diagnostics applied to weather time scales equally exhibit climate model errors. Through data assimilation, these errors can actually be traced back to individual model processes whereby tendencies of key model parameters between analysis cycles are compared to analysis increments, which represent the corrections derived from observations applied to model forecast \cite{Rodwell2007}. Secondly, data assimilation and the wealth of observational information can be used in parameter estimation methods, in which uncertain model parameters and settings become part of the optimal estimation process, that eventually produces the initial conditions but also optimal parameter settings~\cite{Ruiz2015}. Both application areas offer significant potential for weather and climate model development. The adaptation of global data assimilation algorithms to the desirable 100 m - 1 km scales followed by the implementation of both key model error diagnostics and parameter optimisation methods represents another Grand Challenge at present. 

\subsection{{Grand Challenges for high-performance computing}}
In the past, prediction model and data assimilation enhancements have benefited from the exponential growth of computing power \cite{Heath2015}. As this trend is reaching physical limits, entirely new ways of bringing large, compute and data intensive applications onto high-performance infrastructures are needed~\cite{Lawrence2018}. This is the third Grand Challenge.

A generic feature of weather and climate model codes is that they only achieve about~5\% sustained performance on general-purpose processors, mostly because of too much costly data movement\cite{Kogge2013}.

The answer to the computing and data challenge is a combination of doing less, doing it cheaper and doing it with a specific focus on what new processors and system architectures have to offer. This diverse set of solutions requires prediction systems to build in much more flexibility on both sides: the scientific front-end and the computing back-end. 

In terms of numerical methods and model dynamical cores at the front-end, enhanced parallelism means that grid-point models only requiring nearest-neighbour data communication have advantages over the classic, spectral methods that require global communication~\cite{Cote2015} even though the latter still perform very well~\cite{Wedi2014}.

Since performance is mostly limited by memory bandwidth, even higher-order methods have potential today as they deliver more accuracy with invisible computing overheads. However, time stepping is highly relevant because explicit time stepping schemes, which are required for stable calculations with most grid-point models, may avoid global data communication but still imply costly, locally performed data movements; however, much more frequently than semi-implicit or implicit schemes. The ‘impliciteness’ also determines how local or global the solver needs to be, and how well the computations can be parallelised. Advection methods are important in this context as well because they require halo-communication.

Efficiency gains can be obtained from limiting higher resolution to areas of interest \cite{Zaengl2015}, through dynamical grid refinements in areas of dynamic activity and sharp state gradients~\cite{Kuehnlein2012} and by implementing multiple resolution for different prognostic variables. The first option is less suitable for global and longer-range predictions as finer-scale motions would be systematically misrepresented in areas with lower resolution. The second option has significant implications on load-balancing as the computing and communication load across many compute nodes would need to be reassessed and adapted every time step. The third option is a simplified version of the first and offers both flexibility and performance as it trades off resolution against error tolerance at full global scale. For example, while cloud variables need to be updated at every grid point and time step at the highest possible rate, aerosols and most trace gases could be run at coarser scales as they do not vary as much and do not undergo rapid physical and chemical processes. An important ingredient for such front-end flexibility, however, is a data structure that allows flexible mesh and grid handling of all fields, that performs cost effective interpolations and that is fully parallelised~\cite{Deconinck2017}. 

At the computational back-end, an interface to different types of processors is also needed so that memory layout and parallelism can be defined away from the science code. Separating science code from those operations that are hardware dependent is an entirely new concept. While traditional programming models allow shared- and distributed-memory parallelisation at science code level, true flexibility and hardware-portability can only be achieved through this so-called separation of concerns~\cite{Schulthess2015}. 

The re-emergence of artificial intelligence (deep-learning) methods caused by prominent commercial applications and supported by specialised processing technologies also presents potential in Earth-system prediction. Replacing physics based models as a whole may not be possible due to the very large number of degrees of freedom and the strong non-linearity of the system~\cite{Dueben2018}. However, there are successful studies for the prediction of selected parameters at coarse scale~\cite{Nooteboom2018} and short lead times or selected locations~\cite{Baboo2010}, also presenting opportunities for commercial applications.

At model process level, there are benefits for tuning uncertain parameters with better and more comprehensive training, but the key application area for deep-learning methods is to replace or accelerate costly model components. For parameterisations, radiation and cloud schemes are obvious candidates for which good results have been achieved~\cite{Krasnopolsky2006}, however, conservation of mass and energy are important requirements. Going one step further and representing sub grid-scale cloud-dynamics by neural networks that have been trained with three-dimensional large-eddy simulations has been proposed~\cite{Schneider2017} but may be impossible to train for global applications and may require too costly neural networks for capturing the full dimension of the problem.

Lastly, Earth-system model configurations need to be scrutinized depending on the specific application. For example, medium-range weather prediction clearly requires atmosphere-ocean coupling, but does costly, deep ocean circulation matter? How many aerosol prognostic variables need to be included in a weather model compared to an air-quality model? Can time-critical ensembles be run with a pseudo-ensemble in which ensemble spread is calculated by neural networks rather than costly physics based models? What is the best trade-off between spatial resolution - a key factor for physical realism of models (see first challenge) - and model complexity in climate models?

Future models will need to include all such sources of efficiency gains to achieve spatial resolutions that help overcome key sources of model error. Both weather and climate models need the same algorithmic flexibility and generic solutions for software development, even if individual choices about model configuration may differ. The same applies to solutions for handling massive amounts of data to be post-processed, archived and disseminated~\cite{Overpeck2011}. While this aspect is not the subject of this paper, the data challenge is intimately connected to the computing challenge and requires community wide, sustainable solutions. Note that the first two challenges can only be addressed by solving challenge number three – an investment in weather and climate domain specific computational science will therefore be essential to advance predictive skill much further {and therefore help in addressing SDG $\#13$ (Climate Action)}.

\subsection{\modif{Outlook}}
\modif{Advancing weather and climate prediction beyond the present, incremental progress requires a significant investment in all areas of observation and simulation of the fluid envelope of our planet and the oceans as well as their interaction with land surfaces, cryosphere and biosphere. This topic covers scales from meters to thousands of kilometers and processes that operate on time scales between seconds and seasons - even longer if volcanic ash dispersion and glacier/ice sheet processes are included. For the fluid dynamics components, present-day systems make clear choices about which scales to resolve with known equations and which non-resolved scales (and not fully understood processes) to approximate with so-called parametrisations. This choice is largely driven by computational limits, and overcoming these limits through new approaches to designing models, handling large amounts of data and exploiting novel digital technologies is potentially the biggest challenges of this community in the next decade. This has been recognized and is being pursued in several national and international programmes (e.g. US DoE Exascale programme, European Destination Earth action), also promising significant funding and a perspective for addressing SDGs more effectively in the future. }

\section{Conclusion}
\label{Conclusion}

In this paper, we have presented and discussed  a wide range of  Grand  Challenge problems {in Environmental Fluid Mechanics (EFM)} to be tackled as we strive towards a more sustainable planet. They range from fundamental advances in understanding and modeling of stratified turbulence and consequent mixing, to applied studies of pollution transport in the ocean, atmosphere and urban environments. 

{An important consideration in tackling Grand Challenge problems is the juxtaposition of science and engineering}. For example, those developing flow-based solutions in the urban environment are often directed towards building simulations of the system, whereas the scientists are interested in understanding their observations of the natural world through modeling. In both cases, however, the modeling approach embraces the idea of simplification, and the use of dimensional analysis to ensure the models capture the dominant effects. 
Another very important example is the planning of a field experiment to monitor the potential environmental impacts of deep-sea mining, with sea floor surveillance designed to follow any sediment plumes generated by the process. In this context, the engineering process has the potential to disturb a deep-marine habitat and so careful measurement and modeling based on rigorous science is needed to understand the possible impacts. {The key take away is that for grand scale problems such as these, both engineering and scientific approaches are in order}.

{While there is great importance in understanding the fundamentals and using this to build both scaling laws and accurate models, the potential transformation in modelling associated with the advent of large data sets, and the ability to recognise patterns and rules within such data, is also recognized}. This can lead to data based models to complement predictive modeling. It is key, however, to not lose sight of the value of fundamental physics based models in identifying bounds on particular flow regimes, as indicated by dimensionless parameters.  This can help with developing predictive models for highly non-linear processes for which data based models may not always capture such transitions in behaviour.  

In this perspective paper, \modif{given the breadth of the topic}, we have highlighted many \modif{areas in} which EFM plays a central role in sustainability for the planet, and \modif{which in many cases} scientists have been engaged for decades. Continued \modif{and elevated} efforts are needed on all fronts. 

Flood prediction and hurricane forecasting, \modif{for example,} are topics that rely heavily on EFM. These events are increasing in number and intensity, and are  encountered in so many different parts of the worlds nowadays, with increasingly devastating effect, that advances in flood and hurricane prediction are paramount. Similarly, anthropogenic discharges into the environment like the Deepwater Horizon oil spill (the largest marine oil spill in the history of the petroleum industry), discharges from ships (still happening too regularly), or the introduction of plastics of all size into water bodies, increasingly demand a call-to-arms for the EFM community to assess and communicate the extent and scale of their impact.

{Understanding the impact of deep-sea mining is a new arena for EFM, but one of profound importance. Nodules found on the deep seabed contain vast deposits of nickel, cobalt, copper, and manganese, four minerals that are essential for energy storage. As society moves toward driving more electric vehicles and utilizing renewable energy there will be an increased demand for these minerals, to manufacture the batteries necessary to decarbonize the economy. The collection of nodules from the seabed is being considered as a new means for obtaining these materials, but before so doing it is imperative to fully understand the environmental impact of mining resources from the deep ocean and compare it to the environmental impact of mining resources on land. A central question is to understand how the sediment plumes generated by the collection of nodules from the seabed will be carried by water currents~\cite{TomMatthew}.}

\modif{And, of course,} the Covid pandemic has brought to the fore a hugely important topic in EFM for global well being. Key to making progress is for EFM researchers to collaborate with a wide range of experts from other fields to help identify key and relevant problems to inform, and to help assess the impact of EFM research on the problems. Some of the contributions to be made include: (i) the collection of relevant data to describe the environmental system - this can be very difficult, especially in remote or dangerous environments (e.g. the measurement of the dispersal patterns of aerosols produced by breathing or coughing and their sampling in hospital/other settings to assess whether they contribute to infection transmission)~\modif{\cite{Andy1}}; (ii) the design of new experiments (either in the field or the lab) to help
understand the physical processes controlling or influencing the system (e.g. aerosol dispersion patterns)~\modif{\cite{Andy2}}; (iii) the development of simplified physical models that provide a framework to interpret the data can help identify leading order solutions and modifications to help improve the situation or evolve the situation to reduce the undesirable effects (e.g.
increasing ventilation rates in buildings/social distancing); (iv) modeling of the system, perhaps using numerical simulation, as a longer term research endeavor (e.g. the interactions of aerosols with face masks, filters or surfaces, combined with the evaporation processes). In combination, these activities can play a central role in informing public policy and thereby shaping the way that global society should conduct itself in pandemic scenarios.

\modif{
In concluding, given the title of this article and the major contribution the EFM community can make to the challenges facing our planet, it is appropriate to take this opportunity to share some collective thoughts for consideration. We propose the following plan of action for the coming decade. The first action item is to implore researchers to engage with, and commit to using their research to help achieve, the UN Sustainability Development Goals. Ask not what your research can do just for your direct research community, but more broadly identify key sustainability grand challenges, and work with policy makers to support informed decision making up to the highest levels of governance. The second action item is to recognize that tackling the wide array of environmental flow challenges will, more than ever, demand:
\begin{itemize}
\item Advances in theoretical and numerical modeling for atmosphere and ocean flows across the full-range of scales to enhance the quality of model predictions.
\item Advances in experimental techniques and ambitious field measurement campaigns across all scales to 
uncover some of the complex physics of flows, to build and test hypotheses about the controlling processes,  to develop data-driven models and to test theoretical models or numerical simulations.
\end{itemize}
It is then beholden on us as authors to provide some suggestions for key topics to work on in the coming decade, which is our third action item. Our considered opinion is that the following topics (listed in no particular order) are where substantial advances in EFM can make the biggest contribution to society and the health of the planet in the coming decade:  
\begin{itemize}
\item Coastal flood prediction, sea-level rise and its consequences, 
and extreme weather events in the global coastal ocean, to contribute to resilience planning.
\item Collection of pollution transport data and development of new models for dispersion of pollutants  in the atmosphere, oceans, urban environments and groundwater, to support mitigation of disasters and develop long-term solutions.
\item Collection and analysis of data on the dynamics of glaciers and ice sheets, and sea/ice interactions, to 
constrain and build improved models to provide more constraints about  the rate of 
melting ice and its environmental impact.
\item The design of sustainable cities of the future, to enhance human health and well-being in the face of increasing urbanisation of the global population.
\end{itemize}}

\acknowledgments
We are grateful to all the participants of the workshop for their stimulating presentations
and discussion and whose feedback contributed to the contents of this review.
The Les Houches winter school and the present paper were
made possible thanks to the financial support of the following
institutions,  the contribution of which is gratefully
acknowledged: La Formation Permanente du CNRS, Ecole Normale Supérieure de Lyon, IdexLyon and
MIT Environmental Solutions Initiative.
The assistance of the permanent staff of the School, especially M.
Gardette, A. Glomot and I. Lelievre, has been of
invaluable help at every  stage of the preparation and
development of this initiative.

T.D. acknowledges support by the Agence Nationale de la Recherche through Grant ``DisET'' No.~ANR-17-CE30-0003. T.P. acknowledges support by the NSF Hazards-SEES Grant 1520825.
G.H. acknowledges support from the Turbulent Superstructures priority
program of the German National Science Foundation (DFG). N.P. and A.A.S-N acknowledge support from the AtlantOS H2020 project.
{Experiments reported in Fig.~\ref{fig:DryMultiphase}(a) were performed by Ms. Elze Porte, while those in Fig.~\ref{fig:DryMultiphase}(b) by Dr. Amalia Thomas.
Experiments reported in Fig.~\ref{fig:oil-PDF} were performed by Dr. Antonio Augusto Sepp-Neves.}
We are grateful to D. Karrasch for his work on Fig.~\ref{fig:ocean-adv-diff}.
{We thank the United Nations for the permission to include the SDG icons (Fig.~\ref{fig:SDGs}).
The content of this publication has not been approved by the United Nations and does not reflect the views of the United Nations or its officials or Member States.}


\begin{thebibliography}{}

\bibitem{TalksLesHouches}
Invited talks recorded during the conference are available at
{\tt http://perso.ens-lyon.fr/thierry.dauxois/GrandChallenges/speakers.html}.

\bibitem{REFClaudiaAndy}
G. Voet, J. B. Girton, and M. H. Alford,
Pathways, Volume Transport, and Mixing of Abyssal Water in the Samoan Passage,
{\it Journal of Physical Oceanography} 45, 562 
(2015). 

\bibitem{tamay}
G. Novelli,  C. Guigand, C. Cousin, E. Ryan, N. Laxague, H. Dai, B. Haus, and T. \"Ozg\"okmen
A Biodegradable Surface Drifter For Ocean Sampling On A Massive Scale. 
{\it J. Atmos. Oceanic Technol.} 34, 2509 
(2017).

\bibitem{sousa2019}
J. Sousa and C. Gorl\'e, {\it Building and Environment}  154, 13 (2018).

\bibitem{song_etal18}
J. Song, S. Fan, L. Lin,  W.~Mottet, H. Woodward, M.~G. Davies~Wykes, R. Arcucci, D. Xiao, J.-E. Debay, H. ApSimon, E. Aristodemou,
{\it Building Research \& Information} 46, 809 (2018).


\bibitem{Bourouiba1}
{S. Poulain and L. Bourouiba, Biosurfactants change the thinning of contaminated bubbles at bacteria-laden water interfaces. 
{\it Physical Review Letters} 121, 204502 (2018).}

\bibitem{Bourouiba2}
{S. Poulain. and L. Bourouiba,  Disease transmission via drops and bubbles. {\it Physics Today} 72, 70 (2019).}

\bibitem{Mingotti}
{N. Mingotti and A.W. Woods, On the transport of heavy particles through an upward displacement-ventilated space,
{\it Journal fo Fluid Mechanics} {\bf 772}, 478 
(2015).}


\bibitem{Andy1}
\modif{N. Mingotti, D. Grogino, G. Dello Ioio, M. Curran, K. Barbour, C. Howarth, A.  Floto, AW. Woods,  American Journal of Respiratory and Critical Care Medicine (2020).}

\bibitem{Andy2}
\modif{N. Mingotti, R. Wood, C. Noakes, A.W. Woods,  J Fluid Mech. 903, A52 (2020).}

\bibitem{REFCarolinePaul}
P.F. Linden, The fluid mechanics of natural ventilation, {\it Ann. Rev.
Fluid Mech}. {\bf 31}, 201 
(1999).

\bibitem{GladstoneWoods}
{C. Gladstone and A. W.  Woods, 
On buoyancy-driven natural ventilation of a room with a heated floor,
{\it  Journal of Fluid Mechanics} {\bf 441}, 293 
(2001).}

\bibitem{Lohse}
Y. Yang, R.  Verzicco, and D. Lohse,  Scaling laws and flow structures of double diffusive convection in the finger regime,
{\it Journal of Fluid Mechanics} 802,  667 
(2016). 

\bibitem{REFTom}
A. J. Rzeznik, G. Flierl and T. Peacock, 
Model investigations of discharge plumes generated by deep-sea nodule mining operations, 
{\it Ocean Engineering} 172, 684-696. (2019). 

\bibitem{REFDiBenidetto}
M.H. Di Benedetto, J. R. Koseff and N. T. Ouellette, Orientation dynamics of non-spherical particles under surface gravity waves, 
{\it Physical Review Fluids} 4, 034301. (2019)

\bibitem{REFPhilippe} 
M. Robbe-Saule, C. Morize, R. Henaff, Y. Bertho, A. Sauret and P. Gondret, Experimental investigation of tsunami waves generated by granular collapse into water, submitted to {\it Journal of Fluid Mechanics} 907, A11 (2021).

\bibitem{Vowinckel2019}
B.  Vowinckel, J. Withers, P. Luzzatto-Fegiz,  and E.  Meiburg,  Settling of cohesive sediment: particle-resolved simulations, {\it Journal of Fluid Mechanics} 858, 5 (2019).

\bibitem{Dueben2018}
P.D. Dueben and P. Bauer,  Challenges and design choices for global weather and climate models based on machine learning, Geosci. Model Dev. 11, 3999 (2018) 

\bibitem{Johnson1985}
K. L. Johnson,   {\it Contact Mechanics},  Cambridge University Press (1985).

\bibitem{Jop2006}
P.  Jop, Y. Forterre, and O. Pouliquen,   A constitutive law for dense granular flows, {\it Nature} 441, 727 
 (2006). 

\bibitem{Kamrin2012}
K. Kamrin and G. Koval,  Nonlocal Constitutive Relation for Steady Granular Flow, {\it Physical Review Letters} 108, 178301 (2012).

\bibitem{Bouzid2013}
M. Bouzid,  M. Trulsson, P. Claudin , E. Clement, and B. Andreotti, Nonlocal Rheology of Granular Flows across Yield Conditions {\it  Physical Review Letters} 111, 238301 (2013).

\bibitem{Gray2014}
J.M.N.T. Gray and  A.N. Edwards,  A depth-averaged $\mu(I)$-rheology for shallow granular free-surface flows, {\it  Journal of Fluid Mechanics} 755, 503 
 (2014). 

\bibitem{Pouliquen1997}
O. Pouliquen, J.  Delour, and S.B. Savage, Fingering in granular flows, {\it  Nature} 386 (1997).

\bibitem{Hunt2010}
M.L. Hunt and N.M. Vriend,   Booming Sand Dunes, {\it Annual Review of Earth and Planetary Sciences} 38, 281 
 (2010).

\bibitem{VanGerner2007}
H.J. Van Gerner,  M.A. van der Hoef, D.  van der Meer, and K.  van der Weele, Interplay of air and sand: Faraday heaping unravelled, {\it Physical Review E} 76, 051305 (2007). 

\bibitem{Weis2017}
S. Weis  and M. Schroeter,   Analyzing x-ray tomographies of granular packings, {\it  Review of Scientific Instrument} 88, 051809 (2017). 

\bibitem{Parker2017}
D.J. Parker,  Positron emission particle tracking and its application to granular media, {\it  Review of Scientific Instruments} 88, 051803 (2017). 

\bibitem{Thomas2019}
A.L. Thomas and N.M. Vriend,   Photoelastic study of dense granular free-surface flows, {\it Physical Review E} 100, 012902 (2019).  

\bibitem{Thomasetal2019}
A.L. Thomas, Z. Tang, K.E. Daniels, N.M. Vriend, Force fluctuations at the transition between quasistatic to inertial granular flow, Soft Matter 15, 8532 (2019).

\bibitem{Dundas2017}
C.M. Dundas, A.S. McEwen, S. Diniega, C.J. Hansen, S.  Byrne, and J.N. McElwaine,  The formation of gullies on Mars today. In:  S.J. Conway, J.L. Carrivick, P.A., T.  Carling  de Haas and, T.N.
Harrison  (eds) Martian Gullies and their Earth Analogues. Geological Society, London, Special Publications, 467 (2017).

\bibitem{McEwen2007}
A. S. McEwen, E. M. Eliason, J. W. Bergstrom, N. T. Bridges,
C. J. Hansen, W. A. Delamere, J. A. Grant, V. C. Gulick,
K. E. Herkenhoff, L. Keszthelyi, R. L. Kirk, M. T. Mellon,
S. W. Squyres, N. Thomas, and C. M. Weitz,
 Mars Reconnaissance Orbiter’s High Resolution Imaging Science Experiment (HiRISE), {\it  Journal of Geophysical Research} 112, E05S02 (2007).

\bibitem{Meiburg2010}
E.  Meiburg, and B. Kneller,  Turbidity currents and their deposits, {\it  Annu. Rev. Fluid Mech.} 42 (1), 135
 (2010).

\bibitem{Necker2002}
F. Necker,  C. Haertel, L. Kleiser,  and E. Meiburg,   High-resolution simulations of particle-driven gravity currents, {\it Int. J. Multiphase Flow} 28 (2), 279 
 (2002).

\bibitem{Cantero2009}
M.I. Cantero,  S. Balachandar,  A. Cantelli,  C. Pirmez, and G. Parker,   Turbidity current with a roof: Direct numerical simulation of self-stratified turbulent channel flow driven by suspended sediment, {\it J. Geophys. Res.} 114, C03008 (2009).

\bibitem{Yu2014}
X. Yu, T.  Hsu, and S. Balachandar,   Convective instability in sedimentation: 3-d numerical study, {\it  J. Geophys. Res.} 119 (11), 8141 
 (2014).

\bibitem{Burns2015}
P.  Burns and E. Meiburg,   Sediment-laden fresh water above salt water: nonlinear simulations,  {\it J. Fluid Mech.} 762, 156 
 (2015).

\bibitem{Alsinan2017}
A. Alsinan, E.  Meiburg, and P. Garaud,   A settling-driven instability in two-component, stably stratified fluids,  {\it J. Fluid Mech.} 816, 243 
 (2017).

\bibitem{Reali2017}
J.F. Reali, P. Garaud, A.  Alsinan, and E. Meiburg,  Layer formation in sedimentary fingering convection {\it  J. Fluid Mech.} 816, 268 
 (2017).

\bibitem{Richardson1954}
 J.F. Richardson and W.N. Zaki,  The sedimentation of a suspension of uniform spheres under conditions of viscous flow, {\it  Chem. Engng Sci.} 3 (2), 65 
  (1954).

\bibitem{TeSlaa2015}
S. Te Slaa, D. S. van Maren, Q. He,  and J. C. Winterwerp,   Hindered settling of silt, {\it  J. Hydraul. Engng;} 141 (9), 04015020 (2015).

\bibitem{Jenkins2002}
J.T. Jenkins  and C. Zhang, Kinetic theory for identical, frictional, nearly elastic spheres {\it  Phys. Fluids} 14, 1228 (2002).

\bibitem{Boyer2011}
F. Boyer,  E. Guazzelli,  and O.  Pouliquen,  Unifying suspension and granular rheology, {\it Phys. Rev. Lett.} 107, 188301 (2011).

\bibitem{Shields1936}
A. Shields,   Anwendung der Aenhlichkeitsmechanik und Turbulenzforschung auf die Geschiebebewegung , {\it  Mitt. Preuss Versuchsanstalt fur Wasserbau und Schiffbau} 26 (1936).

\bibitem{Garcia1991}
M. H. Garcia and G. Parker, Entrainment of bed sediment into suspension,  {\it J. Hydrol. Eng.} 117(4) (1991).

\bibitem{Frank2015}
D. Frank, D. Foster, I.M. Sou, J. Calantoni,  and  P. Chou,  Lagrangian measurements of incipient motion in oscillatory flows, {\it  J. Geophys. Res. Oceans} 120, 244 
 (2015).

\bibitem{Balachandar2010}
S. Balachandar and J. K. Eaton,   Turbulent dispersed multiphase flow,  {\it Annu. Rev. Fluid Mech.} 42, 111 
 (2010).

\bibitem{Mittal2005}
R.  Mittal and G.  Iaccarino,  Immersed boundary methods, {\it  Annu. Rev. Fluid Mech.} 37, 239 
 (2005).

\bibitem{Biegert2017}
E. Biegert, B. Vowinckel,  and E.  Meiburg,  A collision model for grain-resolving simulations of flows over dense, mobile, polydisperse granular sediment beds,  {\it J. Comput. Phys.} 340, 105 
 (2017).

\bibitem{Shaw2003}
R.A.  Shaw,  Particle-turbulence interactions in atmospheric clouds, {\it  Annu. Rev. Fluid Mech.} 35, 183 
 (2003).

\bibitem{Ouillon2019}
R. Ouillon, N.G.  Lensky, V. Lyakhovsky, A. Arnon, and  E.  Meiburg, Halite precipitation from double-diffusive salt fingers in the Dead Sea: Numerical simulations, {\it  Water Res. Res.} 55, 4252 
 (2019).

\bibitem{Kelley2013}
D.H. Kelley,  and N.T. Ouellette,   Emergent dynamics of laboratory insect swarms, {\it  Sci. Rep.}  3, 1073 (2013). 

\bibitem{Houghton2018}
I.A. Houghton, J.R. Koseff, S.G. Monismith,  and J.O. Dabiri,   Vertically migrating swimmers generate aggregation-scale eddies in a stratified column, {\it  Nature} 556 (7702), 497 
 (2018).
 
\bibitem{Ouillon2019b}
R. Ouillon, I.A. Houghton, J.O. Dabiri, and E. Meiburg, Active Swimmers Interacting with Stratified Fluids during Collective Vertical Migration, submitted to {\it J. Fluid Mech.} (2019).

\bibitem{Ouellette2019}
N.T.  Ouellette,  Flowing crowds, {\it Science} 363 (6422), 27 
 (2019). 

\bibitem{Herminghaus2005}
S. Herminghaus,   Dynamics of wet granular matter, {\it  Adv. in Phys.} 54 (3), 221 
 (2005). 

  \bibitem[Sarkar \& Scotti (2017)]{sarkar:2017}
S. Sarkar  and A. Scotti,    
  {From topographic internal waves to turbulence.} {\it 
    Annu. Rev. Fluid Mech.} {\bf 49}, 195 
     (2017).
  
   \bibitem[Gregg {\it et al.}  (2018)]{gregg:2018}
 M. C. Gregg, E. A. D'Asaro, J. J. Riley,  and E.   Kunze, 
  {Mixing efficiency in the ocean.} {\it
    Annu. Rev. Mar. Sci.} {\bf 10}, 443 
     (2018).
  
\bibitem{hipsey:2020}
\modif{M. R. Hipsey, G. Gideon, G. B. Arhonditsis, C. C. Carey,
J. A. Elliott, M. A. Frassl, J. H. Janse, L. de Mora, B. J. Robson,
 A system of metrics for the assessment and improvement of aquatic
ecosystem models {\it  Environ. Model. Software} 128, 104697 (2020).}

\bibitem[Osborn (1980)]{osborn:1980}
T. R. Osborn, 
  {Estimates of the local-rate of vertical diffusion from dissipation 
    measurements.}
  {\it J. Phys. Oceanogr.} {\bf 10}, 83 
   (1980).

\bibitem[Waterhouse {\it et al.} (2014)]{waterhouse:2014}
  A. F.  Waterhouse,  J. A. MacKinnon, J. D. Nash, M. H. Alford, 
   E. Kunze, H. L. Simmons, K. L. Polzin,  L. C. St Laurent, 
  O. M.  Sun,  R. Pinkel,  L. D. Talley,  C. B. Whalen,  T. N. Huussen,
    G.  S. Carter,   I. Fer, S. Waterman,  A. C. N.  Garabato,
   T. B.  Sanford,  and C. M. Lee, 
  {Global patterns of diapycnal mixing from measurements of the turbulent dissipation rate.}
  {\it J. Phys. Oceanogr.} {\bf 44}, 1854--1872 (2014).
 
 \bibitem[Osborn \& Cox (1972)]{osborn:1972}
T. R Osborn and C. S. Cox,
  {Oceanic fine structure.}
  {\it Geophys. Fluid Dyn.} {\bf 3}, 321 
   (1972).

\bibitem[Salehipour {\it et al.} (2016)]{salehipour:2016}
 H. Salehipour, W. R.  Peltier, C. B. Whalen,  and  J. A. MacKinnon,
  {A new characterization of the turbulent diapycnal diffusivities of mass and momentum in the ocean.} {\it
    Geophys. Res. Lett.} {\bf 43}, 3370 
      (2016).

  \bibitem[Mashayek {\it et al.} (2017)]{mashayek:2017}
A. Mashayek, H. Salehipour, D.  Bouffard, C. P. Caulfield, R.
    Ferrari, M. Nikurashin, W.R. Peltier,  and W. D. Smyth,
  {Efficiency of turbulent mixing in the abyssal ocean circulation.} {\it
    Geophys. Res. Lett.} {\bf 44}, 6296 
     (2017).

 \bibitem[Ivey {\it et al.} (2008)]{ivey:2008}
 G. N.  Ivey, K. B. Winters,  and J. R.  Koseff, 
  {Density stratification, turbulence, but how much mixing?} {\it
    Annu. Rev. Fluid Mech.} {\bf 40}, 169 
    (2008).
  
  \bibitem[Arthur {\it et al.}  (2017)]{arthur:2017}
   R. S. Arthur, S. K.  Venayagamoorthy, J. R. Koseff, and  O. B.   Fringer, 
  {How we compute $N$ matters to estimates of mixing in stratified flows.} {\it
    J. Fluid Mech.} {\bf 831}, R2 (2017).

\bibitem[Smyth {\it et al.} (2001)]{smyth:2001}
W. D. Smyth,  J. N. Moum,  and  D. R. Caldwell, 
  {The efficiency of mixing in turbulent patches: inferences from direct simulations and microstructure observations.} {\it
    J. Phys. Oceanogr.} {\bf 31}, 1969 
     (2001).
     
  \bibitem[Mashayek {\it et al.} (2013)]{mashayek:2013}
A. Mashayek, C. P.  Caulfield, and  W. R. Peltier,
  {Time-dependent, non-monotonic mixing in stratified turbulent shear flows: implications for oceanographic estimates of buoyancy flux.} {\it
    J. Fluid Mech.} {\bf 736}, 570 
    (2013).

  \bibitem[Salehipour \& Peltier (2015)]{salehipour:2015}
  H.  Salehipour  and W. R. Peltier,
  {Diapycnal diffusivity, turbulent {P}randtl number and mixing efficiency in {B}oussinesq stratified turbulence.} {\it
    J. Fluid Mech.} {\bf 775}, 464 
     (2015).

\bibitem[Salehipour {\it et al.} (2016b)]{salehipour:2016b}
 { H. Salehipour, W. R. Peltier,    and C. P. Caulfield,
  {Turbulent mixing due to the Holmboe wave instability at high Reynolds number.} {\it
    J. Fluid Mech.} {\bf 803}, 591 
     (2016).}
     
  \bibitem[Winters {\it et al.} (1995)]{winters:1995}
 K. B. Winters, P. N.  Lombard, J. J. Riley,  and  E. A. D'Asaro,  
  {Available potential energy and mixing in density-stratified fluids,} {\it
    J. Fluid Mech.} {\bf 289}, 115 
     (1995).
 
\bibitem[Ivey {\it et al.} (2018)]{ivey:2018}
 G. N. Ivey, C. E. Bluteau, and N. L. Jones, 
  {Quantifying diapycnal mixing in an energetic ocean.} {\it
    J. Geophys. Res.-Oceans} {\bf 123}, 346 
     (2018).

 \bibitem[Bluteau {\it et al.} (2017)]{bluteau:2017}
  C. E. Bluteau, R. G. Lueck, G. N. Ivey, N. L. Jones, J. W. Book, and A. E. Rice,  
  {Determining mixing rates from concurrent temperature and velocity measurements.} {\it
    J.  Atmosp. Ocean. Tech.} {\bf 34}, 2283 
     (2017).

  \bibitem[Mashayek {\it et al.} (2017b)]{mashayek:2017_kh}
  A. Mashayek, C.P.  Caulfield, and W. R.  Peltier,
  {Role of overturns in optimal mixing in stratified mixing layers.} {\it
    J. Fluid Mech.} {\bf 826}, 522 
    (2017).
    
\bibitem[Salehipour {\it et al.} (2018)]{salehipour:2018}
 H. Salehipour, W. R. Peltier,    and C. P. Caulfield,
  {Self-organized criticality of turbulence in strongly stratified mixing layers.} {\it
    J. Fluid Mech.} {\bf 856}, 228 
     (2018).

   \bibitem[Miles (1961)]{miles:1961}
J. W. Miles, 
  {On the stability of heterogeneous shear flows.} {\it
    J. Fluid Mech.} {\bf 10}, 496 
     (1961).
 
   \bibitem[Howard (1961)]{howard:1961}
L. N. Howard, 
  {Note on a paper of {J}ohn  {W}.} {M}iles,  {\it
    J. Fluid Mech.} {\bf 10}, 509 
     (1961).
 
  \bibitem[Thorpe \& Liu (2009)]{thorpe:2009}
 S. A.  Thorpe  and Z. Liu,  
  {Marginal instability?} {\it
    J. Phys. Oceanogr.} {\bf 39}, 2373 
     (2009).
  \bibitem[Smyth {\it et al.} (2019)]{smyth:2019}
W. D. Smyth, J. D. Nash,  and J. N. Moum, 
  {Self-organized criticality in geophysical turbulence.} {\it
 Sci. Rep.} {\bf 9}, 3747 (2019).

  \bibitem[Ivey and Imberger (1991)]{ivey:1991}
G. N. Ivey,  and   J.  Imberger, 
  {On the nature of turbulence in a stratified fluid. 1. {T}he energetics
  of mixing.} {\it
  J. Phys. Oceanogr.} {\bf 21}, 650 
   (1991).

   \bibitem[Maffioli {\it et al.} (2016)]{maffioli:2016}
A. Maffioli,  G.  Brethouwer, and E.
   Lindborg, 
  {Mixing efficiency  in stratified turbulence.} {\it
    J. Fluid Mech.} {\bf 794}, R3 (2016).

\bibitem[Garanaik \& Venayagamoorthy (2019)]{garanaik:2019}
A.  Garanaik  and S. K.  Venayagamoorthy, 
  {On the inference of the state of turbulence and mixing efficiency
    in stably stratified flows.} {\it
    J. Fluid Mech.} {\bf 867}, 323 
     (2019).
     
\bibitem{gargett:1984}
\modif{A. Gargett, T. Osborn  and P.  Nasmyth, Local isotropy and the decay of turbulence
in a stratified fluid {\it  J. Fluid Mech.} 144, 231--280 (1984).}

     
\bibitem[Shih {\it et al.} (2005)]{shih:2005}
L. H. Shih, J. R. Koseff, G. N. Ivey,  and J. H. Ferziger,  
  {Parameterization of turbulent fluxes and scales using homogeneous sheared stably stratified turbulence simulations.} {\it
    J. Fluid Mech.} {\bf 525}, 193 
     (2005).
     
   \bibitem[Monismith {\it et al.} (2018)]{monismith:2018}
S. G. Monismith, J. R.  Koseff,  and B. L. White, 
  {Mixing efficiency in the presence of stratification: {W}hen is it constant?} {\it
    Geophys. Res. Lett.} {\bf 45}, 5627 
     (2018).

\bibitem[Portwood {\it et al.}  (2019)]{portwood:2019}
G. D. Portwood, S. M  de Bruyn Kops,  and C. P. Caulfield, 
  {Asymptotic dynamics of high dynamic range stratified turbulence.} {\it 
Phys. Rev. Lett.} {\bf 122}, 194504 (2019).
  
\bibitem[Dillon  (1982)]{dillon:1982} T. M. Dillon, {Vertical overturns - a comparison of {T}horpe and {O}zmidov length scales.} {\it
    J. Geophys. Res. -Oceans} {\bf 87}, 9601 
   (1982).
 
  \bibitem[Mater {\it et al.}  (2015)]{mater:2015}
B. D. Mater, S. K.  Venayagamoorthy, L. St Laurent,  and J. N.  Moum, 
  {Biases in {T}horpe-scale estimates of turbulence dissipation. {P}art I: {A}ssessments from large-scale overturns in oceanographic data.} {\it
    J. Phys. Oceanogr.} {\bf 45}, 2497 
     (2015).

 \bibitem[Holmes {\it et al.} (2019)]{holmes:2019}
 { R. M. Holmes, J. D. Zika,  and M. H. England, 
 {Diathermal heat transport in a global ocean model.}
 {\it J. Phys. Oceanogr.} {\bf 49}, 141 
 (2019).}
 
  \bibitem[Large {\it et al.} (1994)]{large:1994}
  {W. G. Large, J. C. Mc{W}illiams, and S. C. Doney,
 {Oceanic vertical mixing: A review and a model with a nonlocal boundary layer parameterization.}
 {\it Rev. Geophys.} {\bf 32}, 363 
 (1994).}


\bibitem[Zaron \& Moum (2009)]{zaron:2009}
 { E. D. Zaron and J. N. Moum,
 {A new look at {R}ichardson number mixing schemes for equatorial ocean modeling.}
 {\it J. Phys. Oceanogr.} {\bf 39}, 2652 
 (2009).}



\bibitem{VIATTE2020}
\modif{C. Viatte, C. Clerbaux, C. Maes et al., 
Air Pollution and Sea Pollution Seen from Space,
{\it Surv Geophys} 41, 1583 
 (2020)}.
 
\bibitem{BRACH2018}
\modif{L. Brach,  P. Deixonne, M.-F. Bernard, E. Durand, M.-C. Desjean, E. Perez, E. van Sebille and A. ter Halle.
Anticyclonic eddies increase accumulation of microplastic in the North Atlantic subtropical gyre,
{\it Marine Pollution Bulletin} 126, 191 
 (2018).}

\bibitem{LEBRETON2012}
\modif{L.C.-M Lebreton, S.D. Greer and J.C. Borrero,
Numerical modelling of floating debris in the world’s oceans,
{\it Marine Pollution Bulletin} 64, 
 653 
 (2012).}


\bibitem{ITOF2019}
\modif{ITOPF. Oil Spill Tanker statistics 2019. http://www.itopf.org/knowledge-resources/data-statistics/statistics/}



\bibitem{FERRARO2007}
\modif{G. Ferraro, A.  Bernardini,  M. David, S. Meyer-Roux, O. Muellenhoff, M. Perkovic,  D.  Tarchi and K. Topouzelis. 
Towards an operational use of space imagery for oil pollution monitoring in the Mediterranean basin: A demonstration in the Adriatic Sea,
{\it Marine Pollution Bulletin} 54 (4),  403 
 (2007).}


\bibitem{weiss08}
J.~B. Weiss and A.~Provenzale.
 {\it Transport and Mixing in Geophysical Flows}.
 Springer, Berlin (2008).

\bibitem{haller13}
G.~Haller and F.J. Beron-Vera,
 Coherent {L}agrangian vortices: The black holes of turbulence,
 {\it Journal of Fluid Mechanics}, 731, R4 (2013).

\bibitem{Gurtin10}
M.E. Gurtin, E.~Fried, and L.~Anand,
 {\it The Mechanics and Thermodynamics of Continua},
 Cambridge University Press, Cambridge (2010).
 
\bibitem{okubo70}
A.~Okubo.
 Horizontal dispersion of floatable trajectories in the vicinity of
  velocity singularities such as convergencies,
 {\it Deep-Sea Res.}, 17, 445 
 (1970).
 
 
\bibitem{weiss91}
J.~Weiss.
 The dynamics of enstrophy transfer in two-dimensional hydrodynamics,
 {\it Physica D}, 48, 273 
  (1991).

\bibitem{drouot76a}
R.~Drouot and M.~Lucius,
 {\it Archiwum Mechaniki Stosowanej}
 282, 923 (1976).

\bibitem{drouot76b}
R.~Drouot and M.~Lucius,
 Approximation du second ordre de la loi de comportement des fluides
  simples. lois classiques d\'eduites de l'introduction d'un nouveau tenseur
  objectif,
 {\it Arch. Mech. Stasow.}, 28, 189 (1976).

\bibitem{astarita79}
G.~Astarita,
 Objective and generally applicable criteria for flow classification,
 {\it Journal of Non-Newtonian Fluid Mechanics} 6, 69 
  (1979).

\bibitem{lugt79}
H.J. Lugt.
 The dilemma of defining a vortex,
 {\it Recent Developments in Theoretical and Experimental Fluid
  Mechanics, U. Muller, K. G. Riesner, and B. Schmidt, Eds.}, 13, 309 
  (1979).
  
\bibitem{haller05}
G.~Haller,
 An objective definition of a vortex,
 {\it Journal of Fluid Mechanics}, 525, 1 
  (2005).

\bibitem{haller15}
G.~Haller,
 Lagrangian coherent structures,
 {\it Annual Review of Fluid Mechanics}, 47, 137 
  (2015).
  
\bibitem{hadjighasem17}
A.~Hadjighasem, M.~Farazmand, D.~Blazevski, G.~Froyland, and G.~Haller,
 A critical comparison of {L}agrangian methods for coherent structure
  detection,
 {\it Chaos}, 27, 053104 (2017).

\bibitem{haller18}
G.~Haller, D.~Karrasch, and F.~Kogelbauer,
 Material barriers to diffusive and stochastic transport,
 {\it Proceedings of the National Academy of Sciences},
  115, 9074 
   (2018).

\bibitem{haller19}
G.~Haller, D.~Karrasch, and F.~Kogelbauer,
 Barriers to the transport of diffusive scalars in compressible flows,
 {\it SIAM Journal of Applied Dynamical Systems}, in press
  (2019).
  
  
\bibitem{HALLER2020}
\modif{G. Haller, S. Katsanoulis, M. Holzner, B. Frohnapfel and D. Gatti,  
Objective barriers to the transport of dynamically active vector fields,
{\it Journal of Fluid Mechanics}, 905, A17  (2020).}


\bibitem{eddies}
W.~Cui, W.~Wang, J.~Zhang, and J.~Yang,
 Multicore structures and the splitting and merging of eddies in
  global oceans from satellite altimeter data,
 {\it Ocean Science}, 15(2), 413 
 (2019).

\bibitem{Pearson-Baylor}
B. Pearson and B. Fox-Kemper,
 Log-normal turbulence dissipation in global ocean models,
 {\it Phys. Rev. Lett.}, 120, 094501 (2018).
 
\bibitem{pierrehumbert94}
R. T. Pierrehumbert,
 Tracer microstructure in the large-eddy dominated regime,
 {\it Chaos, Solitons and Fractals}, 4, 1091 
  (1994).

\bibitem{Prashant2015}
P.~D. Sardeshmukh and C. Penland,
 Understanding the distinctively skewed and heavy tailed character of
  atmospheric and oceanic probability distributions,
 {\it Chaos: An Interdisciplinary Journal of Nonlinear Science},
  25(3), 036410 (2015).

\bibitem{Chu}
P.~C. {Chu},
 Statistical characteristics of the global surface current speeds
  obtained from satellite altimetry and scatterometer data,
 {\it IEEE Journal of Selected Topics in Applied Earth Observations
  and Remote Sensing} 2(1), 27 
  (2009).

\bibitem{Ashkenazy}
Y. Ashkenazy and H. Gildor,
 On the probability and spatial distribution of ocean surface
  currents,
 {\it Journal of Physical Oceanography} 41(12), 2295 
  (2011).

\bibitem{Hu}
Y.~Hu and R.~T. Pierrehumbert,
 The advection--diffusion problem for stratospheric flow. part i:
  Concentration probability distribution function,
 {\it Journal of the Atmospheric Sciences}, 58(12), 1493 
 (2001).

\bibitem{Sepp-Neves}
A. A. Sepp-Neves, N. Pinardi, and A. Navarra,
 A general methodology for beached oil spill hazard mapping,
 {\it Frontiers in Marine Science} {7 (65) (2020)}.

\bibitem{MAXIMENKO}
N. Maximenko, J. Hafner, and P. Niiler,
 Pathways of marine debris derived from trajectories of lagrangian
  drifters,
 {\it Marine Pollution Bulletin}, 65(1), 51 
  (2012).
 At-sea Detection of Derelict Fishing Gear.

\bibitem{LIUBARTSEVA}
S.~Liubartseva, G.~Coppini, R.~Lecci, and E.~Clementi,
 Tracking plastics in the mediterranean: 2d lagrangian model,
 {\it Marine Pollution Bulletin}, 129(1), 151 
 (2018).

\bibitem{LIUBARTSEVA2015}
S.~Liubartseva, M.~De Dominicis, P.~Oddo, G.~Coppini, N.~Pinardi, and
  N.~Greggio,
 Oil spill hazard from dispersal of oil along shipping lanes in the
  southern adriatic and northern ionian seas,
 {\it Marine Pollution Bulletin}, 90(1), 259 
 (2015).

\bibitem{SeppNeves2016}
A.~A. Sepp~Neves, N. Pinardi, and F. Martins,
 It-osra: applying ensemble simulations to estimate the oil spill risk
  associated to operational and accidental oil spills,
 {\it Ocean Dynamics}, 66(8), 939 
  (2016).


\bibitem{LETRAON2019}
\modif{P.Y. Le Traon, A. Reppucci, E. Alvarez Fanjul, L. Aouf, A. Behrens, M. Belmonte, A. Bentamy, L. Bertino, V.E. Brando, M.B. Kreiner, 
M. Benkiran, T. Carval, S.A. Ciliberti, H.  Claustre, E. Clementi, G. Coppini, G. Cossarini, M. De Alfonso Alonso-Muñoyerro, 
A. Delamarche, G. Dibarboure, F. Dinessen, M. Drevillon, Y. Drillet, Y. Faugere, V. Fernández, A. Fleming, M.I. Garcia-Hermosa, 
M.G. Sotillo, G. Garric, F. Gasparin, C. Giordan, M. Gehlen, M.L. Gregoire, S. Guinehut, M. Hamon, C. Harris, F. Hernandez, 
J.B. Hinkler, J. Hoyer, J. Karvonen, S. Kay, R. King, T. Lavergne, B. Lemieux-Dudon, L. Lima, C. Mao, M.J. Martin, S. Masina, 
A. Melet, B. Buongiorno Nardelli, G. Nolan, A. Pascual, J. Pistoia, A. Palazov, J.F. Piolle, M.I. Pujol, A.C. Pequignet, E. Peneva, B. Pérez Gómez, 
L. Petit de la Villeon, N. Pinardi, A. Pisano, S. Pouliquen, R. Reid, E. Remy, R. Santoleri, J. Siddorn, J. She, J. Staneva, A. Stoffelen, 
M. Tonani, L. Vandenbulcke, K. von Schuckmann, G. Volpe, C. Wettre and A. Zacharioudaki,
From Observation to Information and Users: The Copernicus Marine Service Perspective, 
{\it Front. Mar. Sci.} 6, 234 (2019).}



\bibitem{Barker2020}
\modif{C.H. Barker, V.H. Kourafalou, C. Beegle-Krause, M. Boufadel, M.A. Bourassa, S.G. Buschang, Y. Androulidakis, E.P. Chassignet, K.-F. Dagestad, D.G. Danmeier, A.L. Dissanayake, J.A. Galt, G. Jacobs, G. Marcotte, T. Özgökmen, N., Pinardi, R.V. Schiller, S.A. Socolofsky, D. Thrift-Viveros, B. Zelenke, A. Zhang, Y.  Zheng, 
Progress in Operational Modeling in Support of Oil Spill Response. 
{\it J. Mar. Sci. Eng.} 8, 668 (2020).}

\bibitem{GROSSI2020}
\modif{M. D. Grossi, M. Kubat and T. M. Özgökmen,
Predicting particle trajectories in oceanic flows using artificial neural networks,
{\it Ocean Modelling} 156 (2020).}






\expandafter\ifx\csname natexlab\endcsname\relax\def\natexlab#1{#1}\fi
\expandafter\ifx\csname bibnamefont\endcsname\relax
  \def\bibnamefont#1{#1}\fi
\expandafter\ifx\csname bibfnamefont\endcsname\relax
  \def\bibfnamefont#1{#1}\fi
\expandafter\ifx\csname citenamefont\endcsname\relax
  \def\citenamefont#1{#1}\fi
\expandafter\ifx\csname url\endcsname\relax
  \def\url#1{\texttt{#1}}\fi
\expandafter\ifx\csname urlprefix\endcsname\relax\def\urlprefix{URL }\fi
\providecommand{\bibinfo}[2]{#2}
\providecommand{\eprint}[2][]{\url{#2}}

\bibitem[{UN2()}]{UN2019}
\emph{\bibinfo{title}{Goal 11: Sustainable cities and communities}},
  \bibinfo{howpublished}{https://www.undp.org/content/undp/en/home/sustainable-development-goals/goal-11-sustainable-cities-and-communities.html}.


 \bibitem{III2020}
 \modif{Facts + Statistics: U.S. catastrophes,
 https://www.iii.org/fact-statistic/facts-statistics-us-catastrophes.}

\bibitem{IEA2018}
 \modif{The Future of Cooling,
 https://www.iea.org/reports/the-future-of-cooling.}
 
 
 \bibitem{Hajat2015}
 \modif{A. Hajat,  C. Hsia  and M. S. O'Neill, 
Socioeconomic Disparities and Air Pollution Exposure: a Global Review,
 {\it Current Environmental Health Reports} 2
 , 440 
  (2015).}
  
  \bibitem[{\citenamefont{Mochida and Lun}(2008)}]{mochida2008}
\bibinfo{author}{\bibfnamefont{A.}~\bibnamefont{Mochida}} \bibnamefont{and}
  \bibinfo{author}{\bibfnamefont{I.~Y.} \bibnamefont{Lun}},
  \bibinfo{journal}{Journal of Wind Engineering and Industrial Aerodynamics}
  \textbf{\bibinfo{volume}{96}}, \bibinfo{pages}{1498} (\bibinfo{year}{2008}), \bibinfo{note}{4th International Symposium on
  Computational Wind Engineering
}.


\bibitem[{\citenamefont{Montazeri and Blocken}(2013)}]{montazeri2013}
\bibinfo{author}{\bibfnamefont{H.}~\bibnamefont{Montazeri}} \bibnamefont{and}
  \bibinfo{author}{\bibfnamefont{B.}~\bibnamefont{Blocken}},
  \bibinfo{journal}{Building and Environment} \textbf{\bibinfo{volume}{60}},
  \bibinfo{pages}{137} (\bibinfo{year}{2013}).

\bibitem[{\citenamefont{Llaguno-Munitxa
  et~al.}(2017)\citenamefont{Llaguno-Munitxa, Bou-Zeid, and
  Hultmark}}]{llaguno2017}
\bibinfo{author}{\bibfnamefont{M.}~\bibnamefont{Llaguno-Munitxa}},
  \bibinfo{author}{\bibfnamefont{E.}~\bibnamefont{Bou-Zeid}}, \bibnamefont{and}
  \bibinfo{author}{\bibfnamefont{M.}~\bibnamefont{Hultmark}},
  \bibinfo{journal}{Journal of Wind Engineering and Industrial Aerodynamics}
  \textbf{\bibinfo{volume}{165}}, \bibinfo{pages}{115} (\bibinfo{year}{2017}).

\bibitem[{\citenamefont{Baker}(2007)}]{baker2007}
\bibinfo{author}{\bibfnamefont{C.}~\bibnamefont{Baker}},
  \bibinfo{journal}{Journal of Wind Engineering and Industrial Aerodynamics}
  \textbf{\bibinfo{volume}{95}}, \bibinfo{pages}{843} (\bibinfo{year}{2007}).

\bibitem[{\citenamefont{Klein et~al.}(2007)\citenamefont{Klein, Leitl, and
  Schatzmann}}]{klein2007}
\bibinfo{author}{\bibfnamefont{P.}~\bibnamefont{Klein}},
  \bibinfo{author}{\bibfnamefont{B.}~\bibnamefont{Leitl}}, \bibnamefont{and}
  \bibinfo{author}{\bibfnamefont{M.}~\bibnamefont{Schatzmann}},
  \bibinfo{journal}{International Journal of Climatology}
  \textbf{\bibinfo{volume}{27}}, \bibinfo{pages}{1887} (\bibinfo{year}{2007}).

\bibitem[{\citenamefont{Schatzmann and Leitl}(2011)}]{schatzmann2011}
\bibinfo{author}{\bibfnamefont{M.}~\bibnamefont{Schatzmann}} \bibnamefont{and}
  \bibinfo{author}{\bibfnamefont{B.}~\bibnamefont{Leitl}},
  \bibinfo{journal}{Journal of Wind Engineering and Industrial Aerodynamics}
  \textbf{\bibinfo{volume}{99}}, \bibinfo{pages}{169} (\bibinfo{year}{2011}).

\bibitem[{\citenamefont{Mooneghi et~al.}(2016)\citenamefont{Mooneghi, Irwin,
  and Chowdhury}}]{asghari2016}
\bibinfo{author}{\bibfnamefont{M.~A.} \bibnamefont{Mooneghi}},
  \bibinfo{author}{\bibfnamefont{P.}~\bibnamefont{Irwin}}, \bibnamefont{and}
  \bibinfo{author}{\bibfnamefont{A.~G.} \bibnamefont{Chowdhury}},
  \bibinfo{journal}{Journal of Wind Engineering and Industrial Aerodynamics}
  \textbf{\bibinfo{volume}{157}}, \bibinfo{pages}{47} (\bibinfo{year}{2016})
.

\bibitem[{\citenamefont{Harms et~al.}(2011)\citenamefont{Harms, Leitl,
  Schatzmann, and Patnaik}}]{harms2011}
\bibinfo{author}{\bibfnamefont{F.}~\bibnamefont{Harms}},
  \bibinfo{author}{\bibfnamefont{B.}~\bibnamefont{Leitl}},
  \bibinfo{author}{\bibfnamefont{M.}~\bibnamefont{Schatzmann}},
  \bibnamefont{and} \bibinfo{author}{\bibfnamefont{G.}~\bibnamefont{Patnaik}},
  \bibinfo{journal}{Journal of Wind Engineering and Industrial Aerodynamics}
  \textbf{\bibinfo{volume}{99}}, \bibinfo{pages}{289 } (\bibinfo{year}{2011}),
  \bibinfo{note}{5th International
  Symposium on Computational Wind Engineering}.

\bibitem{mochida2011}
{A. Mochida, S. Iizuka, Y. Tominaga and I. Yu-Fat Lun, Up-scaling CWE models to include mesoscale meteorological influences, Journal of Wind Engineering and Industrial Aerodynamics {\bf 99}, 187  
 (2011).} 

\bibitem{yamada2011}
{T. Yamada and K. Koike, Downscaling mesoscale meteorological models for computational wind engineering applications, Journal of Wind Engineering and Industrial Aerodynamics {\bf 99}, 199 
 (2011).} 

\bibitem{chow2018}
{J. Bao, F. Katopodes Chow and K. A. Lundquist, Large-Eddy Simulation over Complex Terrain Using an Improved Immersed Boundary Method in the Weather Research and Forecasting Model, Monthly Weather Review {\bf 146}, 2781
 (2018).}

\bibitem{wyszogrodzki2012} 
{A.A. Wyszogrodzki, S. Miao and F. Chen, Evaluation of the coupling between mesoscale- WRF and LES-EULAG models for simulating fine-scale dispersion, Atmospheric Research {\bf 118}, 324 
 (2012).} 

\bibitem{talbot2012}
{C. Talbot, E. Bou-Zeid and J. Smith, Nested mesoscale large-eddy simulations with WRF: performance in real test cases, Journal of Hydrometeorology {\bf 13}, 1412
 (2012).} 


\bibitem[{\citenamefont{Garc{\'\i}a-S{\'a}nchez
  et~al.}(2018)\citenamefont{Garc{\'\i}a-S{\'a}nchez, van Beeck, and
  Gorl{\'e}}}]{garcia2018}
\bibinfo{author}{\bibfnamefont{C.}~\bibnamefont{Garc{\'\i}a-S{\'a}nchez}},
  \bibinfo{author}{\bibfnamefont{J.}~\bibnamefont{van Beeck}},
  \bibnamefont{and}
  \bibinfo{author}{\bibfnamefont{C.}~\bibnamefont{Gorl{\'e}}},
  \bibinfo{journal}{Building and Environment} \textbf{\bibinfo{volume}{139}},
  \bibinfo{pages}{146} (\bibinfo{year}{2018}).

\bibitem[{\citenamefont{Blocken}(2014)}]{blocken2014}
\bibinfo{author}{\bibfnamefont{B.}~\bibnamefont{Blocken}},
  \bibinfo{journal}{Journal of Wind Engineering and Industrial Aerodynamics}
  \textbf{\bibinfo{volume}{129}}, \bibinfo{pages}{69} (\bibinfo{year}{2014}).

\bibitem[{\citenamefont{Neophytou et~al.}(2011)\citenamefont{Neophytou,
  Gowardhan, and Brown}}]{neophytou2011}
\bibinfo{author}{\bibfnamefont{M.}~\bibnamefont{Neophytou}},
  \bibinfo{author}{\bibfnamefont{A.}~\bibnamefont{Gowardhan}},
  \bibnamefont{and} \bibinfo{author}{\bibfnamefont{M.}~\bibnamefont{Brown}},
  \bibinfo{journal}{Journal of Wind Engineering and Industrial Aerodynamics}
  \textbf{\bibinfo{volume}{99}}, \bibinfo{pages}{357} (\bibinfo{year}{2011}).



\bibitem{peherstorfer2018}
{B. Peherstorfer, K. Willcox and M. Gunzburger. Survey of multifidelity methods in uncertainty propagation, interference and optimization. SIAM Review {\bf 60}, 550 
 (2018).}



\bibitem{duraisamy2019}
{K. Duraisamy, G. Iaccarino and H. Xiao. Turbulence modeling in the age of data. Annual Review of Fluid Mechanics {\bf 51}, 357 
 (2019).} 



\bibitem[{WHO()}]{WHO2018}
\emph{\bibinfo{title}{Ambient (outdoor) air quality and health}},
  \bibinfo{howpublished}{https://www.who.int/news-room/fact-sheets/detail/ambient-(outdoor)-air-quality-and-health}.

\bibitem{OklahomaCity}
K.J. Allwine  and J.E. Flaherty, Joint Urban 2003: Study Overview and Instrument Locations. PNNL-15967. Richland, WA: Pacific Northwest National Laboratory (2006).


\bibitem[{\citenamefont{Garc{\'\i}a-S{\'a}nchez
  et~al.}(2017)\citenamefont{Garc{\'\i}a-S{\'a}nchez, Van~Tendeloo, and
  Gorl{\'e}}}]{garcia2017}
\bibinfo{author}{\bibfnamefont{C.}~\bibnamefont{Garc{\'\i}a-S{\'a}nchez}},
  \bibinfo{author}{\bibfnamefont{G.}~\bibnamefont{Van~Tendeloo}},
  \bibnamefont{and}
  \bibinfo{author}{\bibfnamefont{C.}~\bibnamefont{Gorl{\'e}}},
  \bibinfo{journal}{Atmospheric environment} \textbf{\bibinfo{volume}{161}},
  \bibinfo{pages}{263} (\bibinfo{year}{2017}).
  


\bibitem[{\citenamefont{Garc{\'\i}a-S{\'a}nchez and
  Gorl{\'e}}(2018)}]{garcia2018b}
\bibinfo{author}{\bibfnamefont{C.}~\bibnamefont{Garc{\'\i}a-S{\'a}nchez}}
  \bibnamefont{and}
  \bibinfo{author}{\bibfnamefont{C.}~\bibnamefont{Gorl{\'e}}},
  \bibinfo{journal}{Journal of Wind Engineering and Industrial Aerodynamics}
  \textbf{\bibinfo{volume}{176}}, \bibinfo{pages}{87} (\bibinfo{year}{2018}).

\bibitem[{\citenamefont{Gorl{\'e} et~al.}(2015)\citenamefont{Gorl{\'e},
  Garc{\'\i}a-S{\'a}nchez, and Iaccarino}}]{gorle2015}
\bibinfo{author}{\bibfnamefont{C.}~\bibnamefont{Gorl{\'e}}},
  \bibinfo{author}{\bibfnamefont{C.}~\bibnamefont{Garc{\'\i}a-S{\'a}nchez}},
  \bibnamefont{and}
  \bibinfo{author}{\bibfnamefont{G.}~\bibnamefont{Iaccarino}},
  \bibinfo{journal}{Journal of Wind Engineering and Industrial Aerodynamics}
  \textbf{\bibinfo{volume}{144}}, \bibinfo{pages}{202} (\bibinfo{year}{2015}).



\bibitem{UNISDR2018}
United Nations Office for Disaster Risk Reduction,  Annual Report. United Nations Office for Disaster Risk Reduction (UNISDR), 9-11 Rue de Varembe, 1202 Geneva, Switzerland, 109p. (2018).

\bibitem{Lynch2008}
P. Lynch, The origins of computer weather prediction and climate modelling, J. Comp. Phys. 227(7), 3431 
 (2008).

\bibitem{Haarsma2016}
R.J. Haarsma,  M.J. Roberts, P.L. Vidale, C.A. Senior, A. Bellucci, Q. Bao, P. Chang, S. Corti, N.S. Fučkar, V. Guemas, J. v. Hardenberg, W. Hazeleger, C. Kodama, T. Koenigk, L. Ruby Leung, J. Lu, J.-J. Luo, J. Mao, M.S. Mizielinski, R. Mizuta, P. Nobre, M. Satoh, E. Scoccimarro, T. Semmler, J. Small and J.-S. von Storch, High Resolution Model Intercomparison Project (HighResMIP v1.0) for CMIP6, Geosci. Model Dev. 9, 4185 
 (2016). 

\bibitem{WGNE2018}
WGNE, Research Activities in Atmospheric and Oceanic Modelling (WGNE Blue Book 2018). 

\bibitem{Doblas-Reyes2013}
F.J. Doblas-Reyes,  I. Andreu-Burillo, Y. Chikamoto, J. Garcia-Serrano, V. Guemas, M. Kimoto, T. Mochizuki, L.R L. Rodrigues, and G. J. van Oldenborgh, Initialized near-term regional climate change prediction, Nature Comms. 4, 1715 (2013).

\bibitem{Taylor2012}
K.E. Taylor,  R.J. Stouffer, and G.A. Meehl, An Overview of CMIP5 and the Experiment Design. Bull. Amer. Meteor. Soc. 93, 485 
(2012). 

\bibitem{Palmer2005}
T.N. Palmer,  F.J. Doblas-Reyes, R. Hagedorn, and A. Weisheimer,  Probabilistic prediction of climate using multi-model ensembles: from basics to applications, Philos, Trans. R. Soc. Lond. B Biol. Sci. 360 (1463), 1991 
 (2005). 

\bibitem{Cote2015}
 J. Côté, C. Jablonowski, P. Bauer and N. Wedi,  Numerical Methods of the Atmosphere and Ocean, in Brunet, G., S Jones and P. M. Ruti (Eds.), Seamless Prediction of the Earth System: from Minutes to Months, World Meteorological Organization (WMO) No. 1156, Geneva. 101 
  (2015).

\bibitem{Brown2015}
A. Brown,  M. Miller, A. Beljaars, F. Bouyssel, C. Holloway, and J. Petch,  Challenges for sub-gridscale parameterizations in atmospheric models, in: Brunet, G., S Jones and P. M. Ruti (Eds.), Seamless Prediction of the Earth System: from Minutes to Months, World Meteorological Organization (WMO) No. 1156, Geneva.  171 
 (2015).

\bibitem{Bony2015}
 S. Bony, B. Stevens, D.M.W. Frierson, C. Jakob, M. Kageyama, R. Pincus, T.G. Shepherd, S.C. Sherwood, A.P. Siebesma, A.H. Sobel, M. Watanabe, and M.J. Webb, Clouds, circulation and climate sensitivity,  Nature Geosci., 8, 261 
  (2015).

\bibitem{Bauer2015}
P. Bauer, A. Thorpe, and G. Brunet, The quiet revolution of numerical weather prediction, 
Nature  525,  47 
(2015).

\bibitem{Stevens2013}
B. Stevens  and S. Bony, What Are Climate Models Missing?, Science 340 (6136), 1053 
(2013).

\bibitem{Bellprat2016}
O. Bellprat  and F. Doblas‐Reyes,  Attribution of extreme weather and climate events overestimated by unreliable climate simulations, Geophys. Res. Lett. 43(5), 2158 
 (2016). 

\bibitem{Roberts2018}
M.J. Roberts,  P.L. Vidale, C. Senior, H.T. Hewitt, C. Bates, S. Berthou, P. Chang, H.M. Christensen, S. Danilov, M. Demory, S.M. Griffies, R. Haarsma, T. Jung, G. Martin, S. Minobe, T. Ringler, M. Satoh, R. Schiemann, E. Scoccimarro, G. Stephens, and M.F. Wehner, 2018: The Benefits of Global High Resolution for Climate Simulation: Process Understanding and the Enabling of Stakeholder Decisions at the Regional Scale, Bull. Amer. Meteor. Soc., 99, 2341 
 (2018). 

\bibitem{Prodhomme2016}
C. Prodhomme, L. Batté, F. Massonnet, P. Davini, O. Bellprat, V. Guemas, and F.J. Doblas-Reyes,  Benefits of Increasing the Model Resolution for the Seasonal Forecast Quality in EC-Earth, J. Climate 29, 9141 
 (2016). 

\bibitem{Zappa2013}
G. Zappa,  L. C. Shaffrey, and K. I. Hodges,  The ability of CMIP5 models to simulate North Atlantic extratropical cyclones, J. Climate 26, 5379 
 (2013). 

\bibitem{Jung2012}
T. Jung, , M.J. Miller, T.N. Palmer, P. Towers, N. Wedi, D. Achuthavarier, J.M. Adams, E.L. Altshuler, B.A. Cash, J.L. Kinter, L. Marx, C. Stan, and K.I. Hodges, 
High-Resolution Global Climate Simulations with the ECMWF Model in Project Athena: Experimental Design, Model Climate, and Seasonal Forecast Skill. J. Climate 25, 3155 (2012)

\bibitem{Roberts2015}
M.J. Roberts,  P.L. Vidale, M.S. Mizielinski, M. Demory, R. Schiemann, J. Strachan, K. Hodges, R. Bell, and J. Camp,  Tropical Cyclones in the UPSCALE Ensemble of High-Resolution Global Climate Models, J. Climate 28, 574 
 (2015). 

\bibitem{Bony2005}
 S. Bony and J.L. Dufresne,  Marine boundary layer clouds at the heart of tropical cloud feedback uncertainties in climate models, Geophys. Res. Lett., 32, L20806 (2005).

\bibitem{Wehner2011}
M. Wehner,  L. Oliker, J. Shalf, D. Donofrio, L. Drummond, R. Heikes, S. Kamil, C. Kono, N. Miller, H. Miura, M. Mohiyuddin, D. Randall, and W.‐S. Yang, Hardware/Software Co-design of Global Cloud System Resolving Models, J. Adv. Model. Earth Sys. 3, M1000:22 (2011). 

\bibitem{Schulthess2019}
T.C. Schulthess,  P. Bauer, N. Wedi, O. Fuhrer, T. Hoefler, and C. Schär,  Reflecting on the goal and baseline for exascale computing: A roadmap based on weather and climate simulations, Comp. Sci. Eng. 21(1), 30 
 (2019). 

\bibitem{Dee2014}
D.P. Dee,  M. Balmaseda, G. Balsamo, R. Engelen, A.J. Simmons, and J. Thépaut,  Toward a Consistent Reanalysis of the Climate System, Bull. Amer. Meteor. Soc., 95, 1235 
 (2014). 

\bibitem{Phillips2004}
T.J. Phillips, G.L. Potter, D.L. Williamson, R.T. Cederwall, J.S. Boyle, M. Fiorino, J.J. Hnilo, J.G. Olson, S. Xie, and J.J. Yio, Evaluating parameterizations in General Circulation Models: Climate simulation meets weather prediction, Bull. Amer. Meteor. Soc. 85, 1903 
 (2004). 

\bibitem{Rodwell2007}
M.J.  Rodwell and T.N. Palmer, Using numerical weather prediction to assess climate models, Q. J. Roy. Meteorol. Soc. 133, 129 
 (2007). 

\bibitem{Ruiz2015}
J. Ruiz,  and M. Pulido,  Parameter Estimation Using Ensemble-Based Data Assimilation in the Presence of Model Error, Mon. Wea. Rev. 143, 1568 
 (2015). 

\bibitem{Heath2015}
M.T. Heath,   A tale of two laws, Int. J. High Perf. Comp. Appl. 29(3), 320 
 (2015). 

\bibitem{Lawrence2018}
B.N. Lawrence,  M. Rezny, R. Budich, P. Bauer, J. Behrens, M. Carter, W. Deconinck, R. Ford, C. Maynard, S. Mullerworth, C. Osuna, A. Porter, K. Serradell, S. Valcke, N. Wedi, and S. Wilsonet, Crossing the Chasm: How to develop weather and climate models for next generation computers?, Geosci. Model Dev. 11, 1799 
(2018). 

\bibitem{Kogge2013}
P.  Kogge and J. Shalf,  Exascale Computing Trends: Adjusting to the "New Normal'' for computer architecture, Comp. Sci. Eng. 15, 16 
(2013). 

\bibitem{Wedi2014}
N.P.  Wedi, Increasing horizontal resolution in numerical weather prediction and climate simulations: illusion or panacea?, Phil. Trans. R. Soc. A 372: 20130289 (2014). 

\bibitem{Zaengl2015}
G. Zaengl,  D. Reinert, P. Rípodas and M. Baldauf,  The ICON (ICOsahedral Non-hydrostatic) modelling framework of DWD and MPI-M: Description of the non-hydrostatic dynamical core, Q. J. Roy. Meteorol. Soc. 141B:563 
 (2015).

\bibitem{Kuehnlein2012}
C. Kuehnlein,  P.K. Smolarkiewicz and A. Dörnbrack,  Modelling atmospheric flows with adaptive moving meshes, J. Comp. Phys. 231, 2741 
 (2012).

\bibitem{Deconinck2017}
W. Deconinck,  P. Bauer, M. Diamantakis, M. Hamrud, C. Kühnlein, P. Maciel, G. Mengaldo, T. Quintino, B. Raoult, P.K. Smolarkiewicz, and N.P. Wedi,  Atlas: A library for numerical weather prediction and climate modelling, Comp. Phys. Comm. 220, 188 
 (2017).
\bibitem{Schulthess2015}
T.C.  Schulthess, Programming revisited, Nature Physics 11, 369 
(2015).


\bibitem{Nooteboom2018}
P.D. Nooteboom,  Q. Y. Feng, C. López, E. Hernández-García, and H.A. Dijkstra,  Using network theory and machine learning to predict El Niño, Earth Syst. Dynam. 9, 969 
 (2018). 

\bibitem{Baboo2010}
S.S. Baboo and I.K. Shereef, An efficient weather forecasting system using artificial neural network. Int. J. Environ, Sci. Devel. 1, 321 
 (2010).
 
 \bibitem{Krasnopolsky2006}
V.M.  Krasnopolsky and M.S. Fox-Rabinovitz,  Complex hybrid models combining deterministic and machine learning components for numerical climate modeling and weather prediction, Neural Netw. 19(2), 122 
 (2006). 

\bibitem{Schneider2017}
T. Schneider,  S. Lan, A. Stuart, and J. Teixeira,  Earth System Modeling 2.0: A blueprint for models that learn from observations and targeted high‐resolution simulations, Geophys. Res. Lett. 44, 12396 
 (2017). 

\bibitem{Overpeck2011}
J.T. Overpeck,  G.A. Meehl, S. Bony, and D.R. Easterling, Climate data challenges in the 21st century, Science 331 (6018), 700 
 (2011). 

\bibitem{TomMatthew}
{T. Peacock and M.H. Alford, Is Deep-Sea Mining Worth It?, Scientific American 318, 72 
 (2018).}


\end{thebibliography}
\end{document}